\documentclass[aps,superscriptaddress,showpacs,twocolumn]{revtex4}

\usepackage{mathmacros}	
\usepackage{mathtools}		
\usepackage{stmaryrd}		
\usepackage{csquotes}
\usepackage{graphicx}

\DeclarePairedDelimiter\abs{\lvert}{\rvert}
\DeclarePairedDelimiter\bigabs{\Big\lvert}{\Big\rvert}
\newcommand{\suchthat}{\mid}
\newcommand{\scalprod}[2]{\langle #1 \vert #2\rangle}
\renewcommand{\vec}[1]{\mathbf{#1}}

\newcommand{\domain}{\ensuremath{\mathcal{D}}}
\newcommand{\sphere}{\ensuremath{\mathbb{S}^2}}

\newcommand{\macro}[1]{\ensuremath{\mathcal{#1}}} 
\newcommand{\meanfield}[1]{\ensuremath{\mathscr{#1}}}
\newcommand{\thermo}[1]{\ensuremath{#1}}		       
\newcommand{\partfun}{\ensuremath{\mathcal{Z}}}

\newcommand{\pseudospectrum}{\Sigma}

\begin{document}

\title{Nonlinear energy transfers and phase diagrams for geostrophically balanced rotating--stratified flows}

\author{Corentin Herbert}
\email{cherbert@ucar.edu}
\affiliation{National Center for Atmospheric Research, P.O. Box 3000, Boulder, CO, 80307, USA}

\begin{abstract}
Equilibrium statistical mechanics tools have been developed to obtain indications about the natural tendencies of nonlinear energy transfers in two-dimensional and quasi two-dimensional flows like rotating and stratified flows in geostrophic balance. In this article, we consider a simple model of such flows with a non-trivial vertical structure, namely two-layer quasi-geostrophic flows, which remain amenable to analytical study. We obtain the statistical equilibria of the system in the case of a linear vorticity-stream function relation, build the corresponding phase diagram, and discuss the most probable outcome of nonlinear energy transfers, both on the horizontal and on the vertical, in the presence of stratification and rotation.
\end{abstract}

\pacs{47.27.-i, 47.32.-y, 47.55.Hd, 05.20.Jj}

\maketitle

\section{Introduction}

Turbulent flows are characterized by strong fluctuations of the velocity field. One aspect of their study focuses on the statistics of these fluctuations at the small spatial scales~\cite{MoninBook}. In spite of the fluctuating nature of the flow, coherent structures appear at large scales in many cases. Two-dimensional turbulence is a prototypical situation where large scale coherent structures emerge~\cite{McWilliams1984,Marteau1995}, but this also happens in different contexts, for instance in three-dimensional flows in the presence of rotation~\cite{Mininni2010b,Mininni2011a}. Geophysical flows constitute a vivid illustration of the coexistence between small-scale turbulence and long-lived structures and mean flows up to the planetary scale. 

A first step in the understanding of the emergence of large scale mean flows and coherent structures resides in the notion of \emph{inverse cascade}. It is well known that two-dimensional flows, due to the existence of two quadratic, positive-definite inviscid invariants, the energy and the enstrophy, exhibit a dual cascade scenario~\cite{Kraichnan1967}: the nonlinear interactions tend to push enstrophy downscale while energy is pulled towards the large scales. Inverse cascades have been subsequently investigated in three dimensional flows, mostly in the presence of anisotropic effects which induce a flow behavior which is reminiscent of two-dimensional turbulence, like rotation or stratification. Geophysical flows are strongly dominated by such effects, to the extent that they are well described by the asymptotic regime called \emph{quasi-geostrophy}, corresponding to small Rossby and Froude numbers. In this regime, the flow is quasi two-dimensional, and there are conservation laws analogous to the two-dimensional case, which lead to the existence of an inverse cascade~\cite{Charney1971}. Outside the quasi-geostrophic regime, inverse cascades have been reported in numerical simulations of rotating/stratified flows~\cite{Metais1996,LSmith2002,Kurien2008,Marino2013b}, although the relative roles of balanced and unbalanced motions are not yet completely clear~\cite{Herbert2014b}.
Interestingly, recent research has also reported cases of inverse energy cascade in three-dimensional, helically constrained flows~\cite{Herbert2014a,Zhu2014}, like Beltrami flows in a von K\'arm\'an experiment~\cite{EricHerbert2012} and decimated numerical simulations~\cite{Biferale2012}.

A valuable tool in the study of inverse cascades is statistical mechanics~\cite{Lucarini2013}. Following the pioneering work of Lee~\cite{Lee1952}, Kraichnan~\cite{Kraichnan1967} introduced simple Gibbs ensemble methods for a Galerkin truncation of the inviscid system to obtain the spectra at absolute equilibrium for the energy and enstrophy in a two-dimensional flow. The study of these spectra shows that for some values of the Lagrange parameters associated with energy and enstrophy conservation, the energy is expected to condense at the large scales~\cite{Kraichnan1975,Kraichnan1980}. More recently, Miller~\cite{Miller1990} and Robert and Sommeria~\cite{Robert1991a,Robert1991b} developed a mean-field theory which explains that due to conservation laws, large scale coherent structures emerge as the most probable outcome of turbulent mixing. Their approach even provides a relation (between vorticity and stream function) which characterizes the large scale mean flows eventually produced by this process. The solutions of this equation are called the \emph{equilibrium states} of the system, and constitute a sub-class of the steady-states admitted by the system. The theory therefore predicts which type of coherent structures (e.g. monopoles, dipoles, jets, rings,...) are expected to develop depending only on a few macroscopic parameters, the conserved quantities of the dynamical equations. This information is summarized in the \emph{phase diagram} of the system.

Equilibrium states and phase diagrams have mostly been computed for two-dimensional flows for various domain geometries~\cite{Chavanis1996a,Chen1997,Chavanis1998b,Venaille2009,Herbert2012a}, and quasi two-dimensional models describing the large scales of geophysical flows, dominated by rotation and stratification, but often neglecting the vertical dimension, like in the equivalent barotropic model~\cite{Bouchet2002,Naso2011,Venaille2011a,Herbert2012b}.
Early studies had considered cases of interest for geophysical flows in the framework of the Kraichnan spectrally truncated theory. Salmon, Holloway and Hendershott~\cite{Salmon1976} thus showed that the large scales (larger than the internal radius of deformation) of a two-layer quasi-geostrophic flow tend to behave as a barotropic flow, while the small scales (smaller than the internal radius of deformation) in each layer are statistically uncorrelated. However, their study relies on the presence of a bottom topography, as it is well-known that the spectrally truncated theory does not predict any mean-flow in the absence of bottom topography (in fact, this is an issue of spontaneous symmetry breaking, which vanishes in the mean-field theory). In a similar framework, Merryfield~\cite{Merryfield1998} showed that continuously stratified quasi-geostrophic models confirmed the tendency to barotropization, while improving the representation of the energetics of the system. It is only recently that this thread was taken up in the context of the Miller-Robert-Sommeria (MRS) theory; Venaille~\cite{Venaille2012b} proved that for continuously stratified quasi-geostrophic flows above a bottom topography, large scale flows with a major fraction of the energy trapped near in bottom currents could be attained as statistical equilibria. It was also confirmed that the flow tends to behave as a barotropic flow~\cite{Venaille2012a}, and the role played by the beta effect in this process was investigated.
This study considered realistic cases with prescribed stratification profiles; as a result it is difficult to build a general classification of the equilibrium states of the system. The present paper builds naturally on their work; going back to a two-layer quasi-geostrophic model, we show that it is possible to obtain explicitly the list of all the equilibrium states as a function of the parameters of the model, for the case of a domain with spherical geometry, and discuss the consequences in terms of the horizontal and vertical transfers of energy in geostrophically balanced flows. In section \ref{dyneqinvsec}, we introduce the two-layer quasi-geostrophic model and its dynamical invariants, in section \ref{statmechsec}, we derive and solve the equation characterizing the statistical equilibria using the MRS theory, and in section \ref{stabsec}, we prove the thermodynamic stability of these solutions and finally build the phase diagram for the system. In section \ref{discsec}, we discuss the physical consequences of these results, focusing on the nonlinear transfers of energy in the vertical direction, or in other words, the ability of the fluid to transfer energy from the baroclinic mode to the barotropic mode.

\section{Two-layer quasi-geostrophic flows on a rotating sphere and their dynamical invariants}\label{dyneqinvsec}

We consider a simple model of a stratified, rotating fluid on a domain $\domain$, namely the \emph{two-layer quasi-geostrophic model}. The fluid is made up of two homogeneous layers with respective depth $H_1=\eta H$ and $H_2=(1-\eta)H$, $H$ being the total depth of the fluid. In each layer, the flow is described by the stream function $\psi_i$. Here, we will consider the case of a spherical geometry: $\domain = \sphere$. 

We introduce the \emph{potential vorticity} in each layer by:
\begin{subequations}
\begin{align}
q_1&= - \Delta \psi_1 + \frac{\psi_1-\psi_2}{\eta R^2}+f,\\
q_2&= - \Delta \psi_2 - \frac{\psi_1-\psi_2}{(1-\eta) R^2}+f,
\end{align}
\end{subequations}
where $R$ is the first Rossby deformation radius~\citep{PedloskyGFD,VallisBook}, and $f=2\Omega\cos \theta$ is the Coriolis parameter. Let $F_1=(\eta R^2)^{-1}$ and $F_2=((1-\eta)R^2)^{-1}$. We can suppose from now on that $F_1 \neq 0, F_2 \neq 0$ (i.e. $R< \infty$); otherwise the dynamics in each layer is decoupled from the other layer. To keep things simple, we will not consider the effect of a bottom topography in this study, and we will restrict to the case of layers of equal depth: $\eta=1-\eta=1/2$, or equivalently, $F_1=F_2=F$.
Using Poisson brackets, the dynamics can be recast in the following form:
\begin{subequations}
\begin{align}
\partial_t q_1 + \{ \psi_1, q_1\}&=0,\\
\partial_t q_2 + \{ \psi_2, q_2\}&=0,
\end{align}
\end{subequations}
where we now have
\begin{subequations}
\begin{align}
q_1&= - \Delta \psi_1 + F(\psi_1-\psi_2)+f,\\
q_2&= - \Delta \psi_2 - F (\psi_1-\psi_2)+f.
\end{align}
\end{subequations}
The parameter $F$ measures the strength of the imposed stratification: the larger $F$, the stronger the stratification. Introducing the deformation wavenumber $k_D=2/R$, we also have $F=k_D^2/2$. Strong stratification therefore corresponds to smaller deformation scale.

Note that we are considering the inertial dynamics of the two-layer quasi-geostrophic model. In this context and in the absence of bottom topography (otherwise, one should add a prescribed topography term to the potential vorticity in the bottom layer), the two layers play symmetric roles, since they have the same depth: $\eta=1-\eta=1/2$. Hence, we expect all our results to be invariant under the transformation $1 \leftrightarrow 2$.

We introduce the \emph{barotropic} mode --- which is simply the vertically integrated stream function, also referred to as the \emph{transport stream function}:
\begin{subequations}
\begin{align}
 \psi_t&=\eta \psi_1 + (1-\eta)\psi_2=(\psi_1+\psi_2)/2,
 \intertext{and the \emph{baroclinic} mode:}
 \psi_c&=(\psi_1-\psi_2)/2. 
 \end{align}
 \end{subequations}
 In a similar way, the barotropic and baroclinic potential vorticities are given by, respectively, $q_t = \eta q_1 + (1-\eta)q_2=(q_1+q_2)/2$ and $q_c=(q_1-q_2)/2$. We have the relations
\begin{subequations}
\begin{align}
q_t&=-\Delta \psi_t +f,\\
q_c&=-\Delta \psi_c + \frac{\psi_c} {\eta (1-\eta)R^2}=-\Delta \psi_c + 2F\psi_c.
\end{align}
\end{subequations}
The dynamics of the barotropic and baroclinic modes is given by
\begin{subequations}
\begin{align}
\partial_t q_t + \{ \psi_t, q_t\}+ \{ \psi_c, q_c\}&=0,\\
\partial_t q_c + \{ \psi_t, q_c\} + \{ \psi_c, q_t\}&=0.\label{baroclinicevoleq}
\end{align}
\end{subequations}
Note in particular that only self-interactions of either the barotropic mode or the baroclinic mode contribute to the evolution of the barotropic mode. On the contrary, the evolution of the baroclinic mode depends on interactions between the barotropic and baroclinic mode. Note that when the two layers are not of equal depth ($\eta \neq 1/2$), there is also a non-vanishing term $(1-2\eta)\{\psi_c,q_c\}$ corresponding to self-interactions of the baroclinic mode in the left-hand side of \eqref{baroclinicevoleq}.
The two-layer quasi-geostrophic model can be seen as a minimal model to investigate geostrophically balanced flows with a non-trivial vertical structure: the barotropic mode plays the role of a vertically independent component, while all the vertical variation is contained in the baroclinic mode.

The total energy of the system (normalized by the total height) reads
\begin{subequations}
\begin{align}
\macro{E}[\psi_1,\psi_2]&=\frac{1}{2H} \int_{\domain} d\vec{r} \int_0^H dz (q-f)\psi,\\
&=\frac{1}{2} \int_{\domain} d\vec{r} \left\lbrack \eta(q_1-f)\psi_1 + (1-\eta)(q_2-f)\psi_2 \right\rbrack,\\
&=\frac 1 2 \int_{\domain} \left \lbrack \eta (\nabla \psi_1)^2 + (1-\eta) (\nabla \psi_2)^2+ \frac{(\psi_1-\psi_2)^2}{R^2} \right\rbrack d\vec{r},\\
&=\frac{\macro{K}[\psi_1]+\macro{K}[\psi_2]}{2}+\macro{E}_p\left[\frac{\psi_1-\psi_2}{2}\right],
\end{align}
\end{subequations}
where we have introduced the kinetic energy $\macro{K}[\psi_i]$ in layer $i$: $\macro{K}[\psi_i] = (1/2) \int_\domain (\nabla \psi_i)^2$, and the potential energy $\macro{E}_p[\psi_c] = F \int_\domain \psi_c^2 $, which depends only on the baroclinic part of the flow. The kinetic part can be decomposed in terms of the barotropic kinetic energy $\macro{K}[\psi_t]$ and the baroclinic kinetic energy $\macro{K}[\psi_c]$.
It is easily checked that $\macro{E}[\psi_t,\psi_c]=\macro{K}[\psi_t]+\macro{K}[\psi_c]+\macro{E}_p[\psi_c]$. The total energy $\macro{E}$ is conserved by the dynamics of the flow. In a similar manner, all the \emph{Casimir invariants} for the first layer $\macro{G}_n[q_1]= \int_{\domain} q_1^n d\vec{r}$ and for the second layer $\macro{G}_n[q_2]= \int_{\domain} q_2^n d\vec{r}$ are conserved. The total value of the Casimir invariants for the full system is given by
\begin{align}
\macro{G}_n[q_1,q_2]&= \int_{\domain} \lbrack \eta q_1^n + (1-\eta) q_2^n\rbrack d\vec{r} = \frac{\macro{G}_n[q_1] +\macro{G}_n[q_2]}{2}.
\intertext{On a spherical domain ($\domain=\sphere$), the vertical projection of the total angular momentum}
\macro{L}[q_1,q_2]&=\int_{\domain} (\eta q_1+ (1-\eta) q_2-f)\cos \theta d\vec{r}=\macro{L}[q_t]
\end{align}
is also a conserved quantity. For simplicity, we do not consider here the precession motion of the angular momentum, which leads to the existence of another conserved quantity, the norm of the angular momentum~\cite{Herbert2013b}.
Like the energy, the Casimir invariants and angular momentum are normalized by the total height $H$. Note that the angular momentum conservation law is only a constraint on the barotropic part of the flow.

\section{Statistical equilibrium states}\label{statmechsec}

\subsection{Mean-field Statistical Theory}

We introduce the \emph{mean-field probability distribution}~\cite{Miller1990,Robert1991a,Robert1991b} for vorticity at point $\vec{r}$: $\rho_1(\vec{r},\sigma)$ and $\rho_2(\vec{r},\sigma)$ are the probabilities to observe vorticity $\sigma$ at point $\vec{r}$ respectively in the first layer and in the second layer. The mean-field statistical entropy is given by:
\begin{widetext}
\begin{equation}
\meanfield{S}[\rho_1,\rho_2]=-\int_{-\infty}^{+\infty} d\sigma \int_{\domain} d\vec{r} \Big\lbrack \eta \rho_1(\vec{r},\sigma) \ln \rho_1(\vec{r},\sigma) + (1-\eta)\rho_2(\vec{r},\sigma) \ln \rho_2(\vec{r},\sigma)\Big\rbrack.
\end{equation}
The statistical equilibrium states are defined by the \emph{mean-field variational problem}:
\begin{equation}
\thermo{S}(\thermo{E},\thermo{L},\{\thermo{\Gamma}_n^{(i)}\})=\max_{\rho_1,\rho_2} \Big\lbrace \meanfield{S}[\rho_1,\rho_2] \suchthat \forall i, \meanfield{N}[\rho_i](\vec{r})=1,
\meanfield{E}[\rho_1,\rho_2]=\thermo{E}, \forall n,i, \meanfield{G}_n[\rho_i]=\thermo{\Gamma}_n^{(i)}, \meanfield{L}[\rho_1,\rho_2]=\thermo{L}\Big\rbrace,
\end{equation}
where $\meanfield{N}[\rho](\vec{r})=\int_{-\infty}^{+\infty} \rho(\vec{r},\sigma)d\sigma$ is the normalization constraint enforced at every point of the domain.
The coarse-grained potential vorticity field is then given by the local average with respect to the mean-field probability distributions solution of the variational problem:
\begin{subequations}
\begin{align}
\overline{q_1}(\vec{r}) &= \int_{-\infty}^{+\infty} \sigma \rho_1(\vec{r},\sigma) d\sigma = -\Delta \overline{\psi_1} + F (\overline{\psi_1}-\overline{\psi_2})+f,\\
\overline{q_2}(\vec{r}) &= \int_{-\infty}^{+\infty} \sigma \rho_2(\vec{r},\sigma) d\sigma = -\Delta \overline{\psi_2} + F (\overline{\psi_2}-\overline{\psi_1})+f.
\end{align}
\end{subequations}

As usual, due to the exactness of the mean-field treatment~\cite{Michel1994b,Robert2000,Bouchet2010}, the contributions stemming from energy fluctuations can be discarded. The mean-field expressions for the (fine-grained) conserved quantities are
\begin{subequations}
\begin{align}
\meanfield{E}[\rho_1,\rho_2]&=\frac{1}{2} \int_{\domain} d\vec{r} \Big\lbrack \eta(\overline{q_1}-f)\overline{\psi_1} + (1-\eta)(\overline{q_2}-f)\overline{\psi_2} \Big\rbrack,\\
&=\frac{1}{2} \int_{\domain} d\vec{r} \int_{\R} d\sigma \Big\lbrack \eta(\sigma-f)\overline{\psi_1}\rho_1(\vec{r},\sigma) +(1-\eta)(\sigma-f)\overline{\psi_2} \rho_2(\vec{r},\sigma) \Big\rbrack,\\
\meanfield{G}_n[\rho_i] &= \int_{\domain} d\vec{r} \int_{\R} d\sigma \sigma^n \rho_i(\vec{r},\sigma),\\
\meanfield{L}[\rho_1,\rho_2] &= \int_{\domain} d\vec{r} \int_{\R} d\sigma \Big\lbrack\eta(\sigma-f)\rho_1(\vec{r},\sigma) + (1-\eta)(\sigma-f)\rho_2(\vec{r},\sigma)\Big\rbrack\cos \theta .
\end{align}
\end{subequations}

The critical points of the variational problem must annihilate the first variations of the entropy functional with constraints:
\begin{equation}
\delta \meanfield{S} - \int_{\domain} d\vec{r}\zeta_1(\vec{r}) \delta \meanfield{N}[\rho_1](\vec{r})-\int_{\domain} d\vec{r}\zeta_2(\vec{r}) \delta \meanfield{N}[\rho_2](\vec{r}) -\beta \delta \meanfield{E}[\rho_1,\rho_2]-\sum_{i,k}\alpha_{ik} \delta \meanfield{G}_k[\rho_i]-\mu \delta \meanfield{L}[\rho_1,\rho_2]=0,
\end{equation}
\end{widetext}
where $\zeta_1(\vec{r}),\zeta_2(\vec{r}),\beta,\mu,\alpha_{ik}$ are the Lagrange multipliers associated to the constraints. For statistically independent variations $\delta\rho_1, \delta \rho_2$, we obtain two relations defining the equilibrium Gibbs states, solutions of the mean-field variation problem:
\begin{subequations}
\begin{align}
\rho_1(\vec{r},\sigma)&= \frac 1 {\partfun_1} g_1(\sigma) e^{-(\beta \overline{\psi_1} + \mu \cos \theta)\sigma},\\
\rho_2(\vec{r},\sigma)&= \frac 1 {\partfun_2} g_2(\sigma) e^{-(\beta \overline{\psi_2} + \mu \cos \theta)\sigma},
\end{align}
\end{subequations}
with the partition functions and small-scale vorticity functions
\begin{subequations}
\begin{align}
\partfun_1 &= \int_{-\infty}^{+\infty} g_1(\sigma) e^{-(\beta \overline{\psi_1} + \mu \cos \theta)\sigma} d\sigma,\\
\partfun_2 &= \int_{-\infty}^{+\infty} g_2(\sigma) e^{-(\beta \overline{\psi_2} + \mu \cos \theta)\sigma} d\sigma, \\
g_1(\sigma)&=e^{-\sum_k \alpha_{1k}\sigma^k}, \quad g_2(\sigma)=e^{-\sum_k \alpha_{2k}\sigma^k}.\notag
\end{align}
\end{subequations}

Averaging with respect to the equilibrium probability distributions, we obtain the $q$ -- $\psi$ relations, called \emph{mean-field equations}, describing the coarse-grained equilibrium states (from now on we do not write the horizontal bars):
\begin{subequations}
\begin{align}
q_1&=- \frac 1 { \beta} \frac{\delta \ln \partfun_1}{\delta\psi_1},\\
q_2&=- \frac 1 { \beta} \frac{\delta \ln \partfun_2}{\delta\psi_2}.
\end{align}
\end{subequations}

\subsection{Linear vorticity--stream function relation}

In full generality, the function relating the coarse-grained vorticity $q_i$ to the stream function $\psi_i$ can be any function. An interesting case, amenable to analytic studies, is the case where this function is linear. This amounts to studying a subset of solutions of the original mean-field variational problem~\cite{Bouchet2008}. This case is also obtained in certain physical circumstances, like the strong mixing limit~\cite{Chavanis1996a} or by choosing a Gaussian prior for the small-scale vorticity~\cite{Chavanis2008b} in the Ellis-Haven-Turkington interpretation of the small-scale vorticity function~\cite{Turkington1999,Ellis2000,Ellis2002}. A linear $q$ -- $\psi$ relation is also obtained when taking into account only the energy, the circulation and the enstrophy as conserved quantities in the variational problem~\cite{Naso2010a}. In the case of barotropic flow on a spherical domain, the phase diagram of the system has been previously built precisely in this context (\cite{Herbert2012a}, see also~\cite{Herbert2012b} for a comprehensive discussion). We will thus restrict here to the same situation, and follow the method introduced by Chavanis and Sommeria~\cite{Chavanis1996a} to solve the linear vorticity-stream function relation.

The $q$ -- $\psi$ relations therefore become (see \cite{Venaille2012a})
\begin{subequations}
\begin{align}
q_1&=-\beta \thermo{Z}_1 \psi_1-\mu \thermo{Z}_1 \cos \theta,\\
q_2&=-\beta \thermo{Z}_2 \psi_2-\mu \thermo{Z}_2 \cos \theta.
\end{align}
\end{subequations}
These relations involve the initial values of the fine-grained enstrophy in each layer $\thermo{Z}_1,\thermo{Z}_2$, and the Lagrange parameters associated with conservation of energy $\beta$ and angular momentum $\mu$. They correspond to the class of equilibrium states for which the fine-grained enstrophy in each layer is much larger than the enstrophy of the coarse-grained field. Note that, as we will not be considering other Casimir invariants than the enstrophy in the sequel, we have changed the notation from $\thermo{\Gamma}_2^{(i)}$ to $\thermo{Z}_i$, for convenience.
We have fixed the \emph{gauge condition} $\langle \psi_1\rangle = \langle \psi_2\rangle =0$. Replacing $q_1$ and $q_2$, we obtain the mean-field equations:
\begin{subequations}
\begin{align}
\Delta \psi_1 - \beta \thermo{Z}_1 \psi_1 -F(\psi_1-\psi_2)&=f+\mu \thermo{Z}_1 \cos \theta,\\
\Delta \psi_2 - \beta \thermo{Z}_2 \psi_2 -F(\psi_2-\psi_1)&=f+\mu \thermo{Z}_2 \cos \theta.
\end{align}
\end{subequations}

Formally, this system has the form of a linear operator $\mathcal{A}_\beta$, acting on the cartesian product $L^2(\domain)\times L^2(\domain)$: $\mathcal{A}_\beta \Psi = \mathcal{B}_\mu$, with
\begin{align}
\mathcal{A}_\beta  \Psi &= \begin{pmatrix} \Delta \psi_1 -  \beta \thermo{Z}_1 \psi_1 -F(\psi_1-\psi_2) \\
\Delta \psi_2 -  \beta \thermo{Z}_2 \psi_2 -F(\psi_2-\psi_1) \end{pmatrix},\\
\Psi&= \begin{pmatrix}\psi_1 \\ \psi_2\end{pmatrix}, \\
\mathcal{B}_\mu &= \begin{pmatrix} f+\mu \thermo{Z}_1\cos \theta \\ f+\mu \thermo{Z}_2\cos \theta\end{pmatrix}.
\end{align}

Solving the system  amounts to inverting this linear operator. Here the invertibility condition is not as clear as in the one-layer model~\cite{Herbert2012a}. To simplify the computations, we will decompose the operator in spectral space: let
\begin{equation}
\psi_1 = \sum_{n=1}^{+\infty} \sum_{m=-n}^n \psi_{nm}^{(1)} Y_{nm}, \quad \psi_2 = \sum_{n=1}^{+\infty} \sum_{m=-n}^n \psi_{nm}^{(2)} Y_{nm}.
\end{equation}
The spherical harmonics $Y_{nm}$ are eigenfunctions of the Laplacian with eigenvalues $-\beta_n$: $\Delta Y_{nm} = - \beta_n Y_{nm}$, with $\beta_n=n(n+1)$.
The vorticity fields can be decomposed in a similar manner, with the following relations between the coefficients in spectral space:
\begin{subequations}
\begin{align}\label{vortcoeff}
q_{nm}^{(1)} & = \beta_n \psi_{nm}^{(1)} +F (\psi_{nm}^{(1)}-\psi_{nm}^{(2)} ) + \delta_{n1}\delta_{m0}f_{10},\\
q_{nm}^{(2)} & = \beta_n \psi_{nm}^{(2)} +F (\psi_{nm}^{(2)}-\psi_{nm}^{(1)} ) + \delta_{n1}\delta_{m0}f_{10},
\end{align}
\end{subequations}
with $f_{10}=2\Omega\sqrt{4\pi/3}$.
We introduce the matrices
\begin{align}
A_n(\beta)&=
\begin{pmatrix} 
\beta_n+ \beta \thermo{Z}_1 +F & -F\\
-F & \beta_n +\beta \thermo{Z}_2 +F
\end{pmatrix},\\
&=\beta_n I_2 + \beta \begin{pmatrix}\thermo{Z}_1 & 0 \\ 0 & \thermo{Z}_2 \end{pmatrix} + F \begin{pmatrix} 1 & -1 \\ -1 & 1\end{pmatrix},
\end{align}
so that
\begin{equation}
\mathcal{A}_\beta \Psi = \mathcal{B}_\mu \iff \left\lbrace
\begin{tabular}{lr}
$A_1(\beta) \Psi_{10} = -\sqrt{\frac {4\pi} 3} \begin{pmatrix} 2\Omega+\mu \thermo{Z}_1 \\ 2\Omega+\mu \thermo{Z}_2\end{pmatrix}$ & \\
$A_n(\beta) \Psi_{nm} = 0, \quad (n,m) \neq (1,0)$
\end{tabular}
\right.,
\end{equation}
with $\Psi_{nm}=\begin{pmatrix} \psi_{nm}^{(1)} \\ \psi_{nm}^{(2)}\end{pmatrix}$. 
The determinant of a $A_n(\beta)$ matrix is given by
\begin{align}
d_n(\beta) &= \det A_n(\beta),\\
&= \thermo{Z}_1\thermo{Z}_2\beta^2 + (\thermo{Z}_1+\thermo{Z}_2)(\beta_n + F)\beta +\beta_n (\beta_n+2F).
\end{align}
It is a second order polynomial in $\beta$, let $\gamma_n^{\pm}$ be its roots. In general, we have:
\begin{equation}
\begin{split}
\gamma_n^\pm = &-\frac{(\thermo{Z}_1+\thermo{Z}_2)(\beta_n+ F)}{2\thermo{Z}_1\thermo{Z}_2} \\
&\pm \frac{\sqrt{ (\thermo{Z}_1-\thermo{Z}_2)^2(\beta_n+F)^2+4F^2\thermo{Z}_1\thermo{Z}_2}}{2\thermo{Z}_1\thermo{Z}_2}.
\end{split}
\end{equation}
When $R \to +\infty$ ($F \to 0$), we recover the one layer case condition: the determinant of $A_n(\beta)$ is simply $d_n(\beta)=(\thermo{Z}_1\beta+\beta_n)(\thermo{Z}_2\beta+\beta_n)$, so that $\gamma_n^\pm$ are just $-\beta_n/\thermo{Z}_1,-\beta_n/\thermo{Z}_2$. In particular, they coincide when $\thermo{Z}_1=\thermo{Z}_2$, so that the operator is not invertible when $\beta \in \Sp \Delta$ ($\thermo{Z}$ being absorbed in the definition of $\beta$ in this case).

\begin{figure*}
\includegraphics[width=\linewidth]{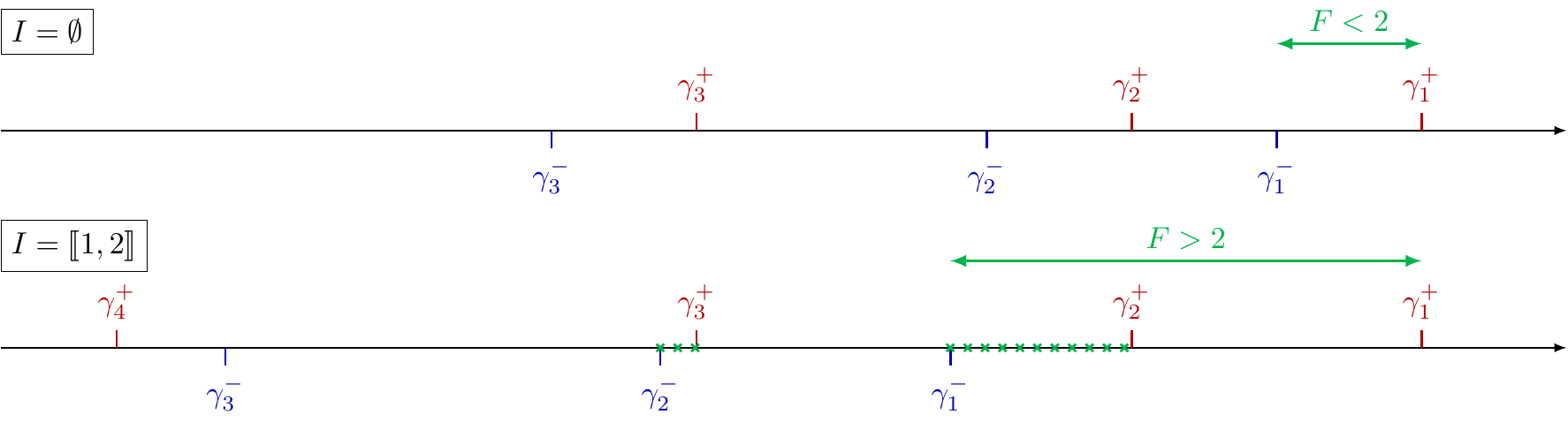}
\caption{(Color online) The roots $\gamma_n^{\pm}$ of the determinant of the linear operator $\mathcal{A}_\beta$ in the case $Z_1=Z_2$. In the first case (top; corresponding to $F=1$ and $\thermo{Z}=1.2$), these roots are not intertwined: $\cdots < \gamma_{n+1}^- < \gamma_{n+1}^+ < \gamma_n^- < \gamma_n^+ < \cdots < \gamma_1^- < \gamma_1^+$ and $I=\emptyset$. In the second case (bottom; corresponding to $F=3.25$ and $\thermo{Z}=1.2$), we have $\gamma_1^- < \gamma_2^+$ and $\gamma_2^- < \gamma_3^+$, so that $I=\llbracket 1, 2\rrbracket$.}\label{gammafig}
\end{figure*}

\begin{figure*}
\includegraphics[width=\linewidth]{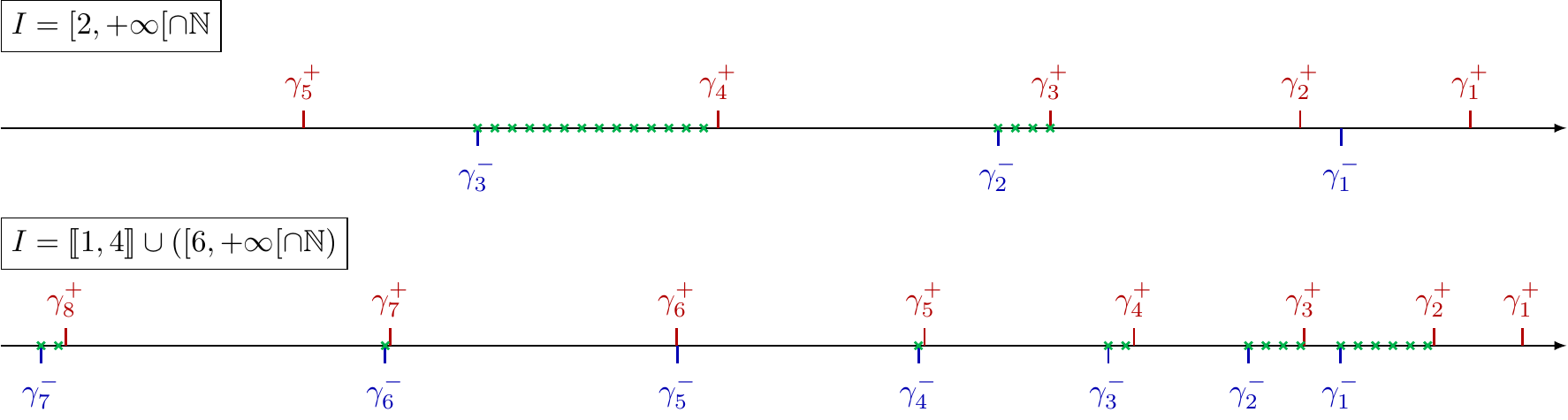}
\caption{(Color online) The roots $\gamma_n^{\pm}$ of the determinant of the linear operator $\mathcal{A}_\beta$ in the case $\thermo{Z}_1 \neq \thermo{Z}_2$. In the first case (top; $F=0.5, \thermo{Z}_1=1, \thermo{Z}_2=2.1$), there is root intertwining after a given rank; in this example, $I=[ 2, +\infty[ \cap \N$. In the second case (bottom; $F=4, \thermo{Z}_1=0.4,\thermo{Z}_2=0.5$), all roots are intertwined except for a finite number of them. Here, $I=\llbracket 1, 4 \rrbracket \cup ([6,+\infty[ \cap \N)$. The $x$-axis has been rescaled in the bottom figure.}\label{gamma2fig}
\end{figure*}

In the two layer case, we need to consider the set $\pseudospectrum=\{ \gamma_n^{\pm}, n \in \mathbb{N}^*\}$, which plays an analogous role to the spectrum of the Laplacian in the one-layer case. Albeit we know a lot about the spectrum of the Laplacian, we know much less in general about the set $\pseudospectrum$. Still, the sequences $(\gamma_n^-)_{n\geq 1}$ and $(\gamma_n^+)_{n\geq 1}$ are both monotonically decreasing ($\gamma_{n+1}^-\leq \gamma_n^-$ and $\gamma_{n+1}^+\leq \gamma_n^+$), tend to negative infinity as $n \to +\infty$, and we always have $\gamma_n^- \leq \gamma_n^+ \leq \gamma_1^+$. Nevertheless, it may happen that for some $n$, $\gamma_n^-< \gamma_{n+1}^+$. When this happens, we say that the roots are \emph{intertwined}; we introduce the set $I=\{ n \in \N, \gamma_{n+1}^+-\gamma_n^->0\}$. When the roots are not intertwined, $I=\emptyset$ and we have:
\begin{equation*}
\cdots < \gamma_{n+1}^- < \gamma_{n+1}^+ < \gamma_n^- < \gamma_n^+ < \cdots < \gamma_1^- < \gamma_1^+. 
\end{equation*}
After some easy but tedious algebra, one obtains a classification of all the possible cases:
\begin{itemize}
\item Let us first assume that $Z_1=Z_2=Z$. Then $\gamma_n^+=-\beta_n/\thermo{Z}$, and $\gamma_n^-=-(\beta_n+2F)/\thermo{Z}$. There are two possibilities (illustrated in Fig. \ref{gammafig}):
\begin{enumerate}
\item $F<2$, there is no root intertwining: $I=\emptyset$.
\item $F\geq 2$, there is root intertwining. In this case, the set $I$ is of the form $I=\llbracket 1,p \rrbracket$ for some $p \in \N^*$.
\end{enumerate}
\item Let us now assume that $Z_1 \neq Z_2$. Then there is always root intertwining, at least for $n$ large enough. There are two cases (illustrated in Fig. \ref{gamma2fig}):
\begin{enumerate}\setcounter{enumi}{2}
\item $I = [p, +\infty [ \cap \N$ for some $p \in \N^*$.
\item $I = \llbracket 1, p \rrbracket \cup ([q, +\infty[ \cap \N)$ for some $p<q \in \N^*$.
\end{enumerate}
\end{itemize}
Note that $I$ is always either an interval of $\N$ or the complement of an interval.
To avoid unnecessary technical complications, we shall always suppose in the sequel that all the elements of $\pseudospectrum$ are pairwise distinct.

\subsection{Solutions of the linear mean-field equation}\label{meanfieldeqsolsec}

\subsubsection{Continuum solution}

When $\beta \notin \pseudospectrum$, all the $A_n(\beta)$ matrices are invertible and the statistical equilibrium state is given by
\begin{align}
\Psi_{nm}&=0, (n,m) \neq (1,0), \\
\Psi_{10} &= -\sqrt{\frac {4\pi} 3 } A_1(\beta)^{-1} \begin{pmatrix} 2\Omega+\mu \thermo{Z}_1 \\ 2\Omega+\mu \thermo{Z}_2\end{pmatrix}.
\end{align}
Finally,
\begin{subequations}
\begin{align}
\psi_1 &= \Omega_1 \cos \theta,\\
\psi_2 &=\Omega_2 \cos \theta,
\end{align}
with
\begin{align}
\Omega_1&=-\frac{1}{d_1(\beta)} \Big\lbrack2\Omega(2F+\beta_1+\beta \thermo{Z}_2)\notag\\
&\phantom{-\frac{1}{d_1(\beta)} \Big\lbrack}+\mu(F(\thermo{Z}_1 +\thermo{Z}_2 ) + \beta \thermo{Z}_1 \thermo{Z}_2+\beta_1 \thermo{Z}_1)\Big\rbrack,\label{omegaSB1eq}\\
\Omega_2&=-\frac{1}{d_1(\beta)} \Big\lbrack2\Omega(2F+\beta_1+\beta \thermo{Z}_1)\notag\\
&\phantom{-\frac{1}{d_1(\beta)} \Big\lbrack}+\mu(F(\thermo{Z}_1 +\thermo{Z}_2 ) + \beta \thermo{Z}_1 \thermo{Z}_2+\beta_1 \thermo{Z}_2)\Big\rbrack.\label{omegaSB2eq}
\end{align}
\end{subequations}
Hence, for the continuum solution, the critical points in the energy-enstrophy case correspond to solid-body rotation in each layer, with different angular velocities. 
The barotropic and baroclinic components are given by:
\begin{subequations}
\begin{align}
\psi_t &= \Omega_t \cos \theta,\\
\psi_c &= \Omega_c \cos \theta,
\end{align}
with
\begin{align}
\Omega_t &= - \frac{1}{d_1(\beta)} \Big\lbrack 2\Omega(2F+\beta_1+\beta [\thermo{Z}_2+\thermo{Z}_1]/2)\notag\\
&\phantom{= - \frac{1}{d_1(\beta)} \Big\lbrack}+\mu(F(\thermo{Z}_1 +\thermo{Z}_2 ) + \beta \thermo{Z}_1 \thermo{Z}_2+\beta_1 [ \thermo{Z}_1+\thermo{Z}_2]/2)\Big\rbrack,\\
\Omega_c &= (\thermo{Z}_1-\thermo{Z}_2) \frac{2\Omega\beta-\mu\beta_1}{2d_1(\beta)}.
\end{align}
\end{subequations}

The angular momentum and energy are given by
\begin{align}
\macro{L}&=\frac{\beta_1}{3}\Omega_t,\\
\macro{E}&= \frac{\beta_1}{6}\Omega_t^2+(\beta_1+2F)\frac{\Omega_c^2}{6},\\
&=\frac{3}{2\beta_1}\macro{L}^2+\frac{1}{8}(\beta_1+2F)(\thermo{Z}_1-\thermo{Z}_2)^2 \frac{(2\Omega\beta-\mu\beta_1)^2}{d_1(\beta)^2}.
\end{align}

Note that in the case of a vertically uniform enstrophy profile ($\thermo{Z}_1 = \thermo{Z}_2$), the continuum solution is just a uniform solid-body rotation: $\Omega_1=\Omega_2$. Then the flow is purely barotropic: $\psi_c = 0$, and $\Omega_t = - \frac{2\Omega+\mu Z}{\beta_1+\beta Z}$.

In general ($Z_1 \neq Z_2$), barotropic flows are obtained when the Lagrange parameters are linked by the relation $2\Omega\beta=\mu\beta_1$. On the contrary, purely baroclinic flows correspond to the case when $\mu=\mu^*(\beta)$, with
\begin{align}
\mu^*(\beta)&=-2\Omega \frac{\beta_1+2F+\beta [ \thermo{Z}_2+\thermo{Z}_1]/2}{F(\thermo{Z}_1 +\thermo{Z}_2 ) + \beta \thermo{Z}_1 \thermo{Z}_2+\beta_1 [ \thermo{Z}_1+\thermo{Z}_2]/2}.
\end{align}
When $Z_1=Z_2=Z$, this reduces to $\mu^*(\beta)=-2\Omega/Z$, and the fluid reaches a state of rest.

\subsubsection{Degenerate cases: $\beta = \gamma_{n>1}^{\pm}$}

Suppose now that $\beta = \gamma_n^{\pm} \in \pseudospectrum$, with $n>1$. Then for each $-n \leq m \leq n$, $\Psi_{nm} \in \Ker A_n(\beta)$, which can be expressed as 
\begin{equation}
\psi_{nm}^{(1)}= \frac{F}{\beta_n+ \thermo{Z}_1\gamma_n^{\pm}+F}  \psi_{nm}^{(2)} = \left(\frac{F}{\beta_n+ \thermo{Z}_2 \gamma_n^{\pm}+F}\right)^{-1}  \psi_{nm}^{(2)},
\end{equation}
given that neither denominators vanish. Finally, the critical points in this case are superpositions of a solid-body rotation and a multipole in each layer: 
\begin{subequations}
\begin{align}
\psi_1 &= \Omega_1 \cos \theta +  \frac{F}{\beta_n+ \thermo{Z}_1\gamma_n^{\pm}+F}  \sum_{m=-n}^n \psi_{nm}^{(2)} Y_{nm}(\theta,\phi),\\
\psi_2 &=\Omega_2 \cos \theta + \sum_{m=-n}^n   \psi_{nm}^{(2)} Y_{nm}(\theta,\phi),
\end{align}
\end{subequations}
where $\Omega_1,\Omega_2$ are given by equations~\eqref{omegaSB1eq}-\eqref{omegaSB2eq} and the $\psi_{nm}^{(2)}$ are arbitrary coefficients satisfying only the reality constraint ${\psi_{nm}^{(2)}}^*=(-1)^m\psi_{n,-m}^{(2)}$ and the total energy constraint:
\begin{align}
\begin{split}
\macro{E}&=  \frac{\beta_1}{6}\Omega_t^2+(\beta_1+2F)\frac{\Omega_c^2}{6}+ \frac{1}{4} \Big\lbrack  \beta_n \abs[\Big]{\frac{F}{\beta_n+ \thermo{Z}_1\gamma_n^{\pm}+F}}^2 \\
&+ \beta_n  + F \abs[\Big]{\frac{F}{\beta_n+ \thermo{Z}_1\gamma_n^{\pm}+F}-1}^2\Big\rbrack \sum_{m=-n}^n \abs{\psi_{nm}^{(2)}}^2,
\end{split}\\
\intertext{When $\thermo{Z}_1=\thermo{Z}_2$, the energy reduces to}
\macro{E}&=\begin{cases} 
\displaystyle\frac{3}{2\beta_1}\macro{L}^2 + \frac{\beta_n}{2}\sum_{m=-n}^n \abs{\psi_{nm}^{(2)}}^2 & \text{ if } \beta=\gamma_n^+\\
\displaystyle\frac{3}{2\beta_1}\macro{L}^2 + \frac{\beta_n+2F}{2} \sum_{m=-n}^n \abs{\psi_{nm}^{(2)}}^2 & \text{ if } \beta=\gamma_n^-
\end{cases}
\end{align}
Note that in this case, when $\beta=\gamma_n^+$, the critical points of the variational problem are purely barotropic flows: $\psi_1=\psi_2=\psi_t$ and $\psi_c=0$, while when $\beta=\gamma_n^-$, the critical points are the sum of a barotropic solid-body rotation and a baroclinic multipole flow: $\psi_t=\Omega_t \cos \theta, \psi_c \neq 0$.
In general, the barotropic and baroclinic modes are given by
\begin{align}
\psi_t &= \Omega_t \cos \theta + \frac{F+(\beta_n+Z_1\gamma_n^{\pm})/2}{F+\beta_n+ \thermo{Z}_1\gamma_n^{\pm}} \sum_{m=-n}^n \psi_{nm}^{(2)} Y_{nm}(\theta,\phi),\\
\psi_c &=\Omega_c \cos \theta - \frac{\beta_n+Z_1\gamma_n^{\pm}}{F+\beta_n+ \thermo{Z}_1\gamma_n^{\pm}} \sum_{m=-n}^n \psi_{nm}^{(2)} Y_{nm}(\theta,\phi),
\end{align}
so that the condition for purely barotropic flow reads 
\begin{equation}
\mu\beta_1=2\Omega\gamma_n^\pm \text{ (or $Z_1=Z_2$) and } \beta_n+Z_1\gamma_n^\pm=0, 
\end{equation}
which does not occur in general (but it is in particular the case when $Z_1=Z_2$ for $\gamma_n^+$). Conversely, the condition for purely baroclinic flows is 
\begin{equation}
\mu=\mu^*(\gamma_n^\pm) \text{ and } Z_1\gamma_n^\pm=-(\beta_n+2F),
\end{equation}
which again does not occur in general but does so in particular when $Z_1=Z_2$ for $\gamma_n^-$.

\subsubsection{Degenerate cases: $\beta = \gamma_1^{\pm}$}

Let us finally treat the case $\beta = \gamma_1^{\pm}$. Solutions of the mean-field equation exist only when 
\begin{equation}
\begin{pmatrix} 2\Omega+\mu \thermo{Z}_1 \\ 2\Omega+\mu \thermo{Z}_2\end{pmatrix} \in \Im A_1(\gamma_1^{\pm}), 
\end{equation}
and it is easily proved that 
\begin{equation}
\Im A_1(\gamma_1^{\pm}) = \R \begin{pmatrix} - F \\ \beta_1+\gamma_1^{\pm} \thermo{Z}_2+F \end{pmatrix}, 
\end{equation}
so that solutions exist only when $\mu=\mu_c^{\pm}$, with
\begin{align}
\mu_c^{\pm} &= -2\Omega\frac{\beta_1+2F+\gamma_1^{\pm} \thermo{Z}_2}{\beta_1\thermo{Z}_1+\gamma_1^{\pm} \thermo{Z}_1\thermo{Z}_2+F(\thermo{Z}_1+\thermo{Z}_2)}.
\end{align}
Then computations similar to the previous case yield
\begin{subequations}
\begin{align}
\psi_1 &=\Omega_1 \cos \theta +  \frac{F}{\beta_1+ \thermo{Z}_1\gamma_1^{\pm}+F} a \sin \theta \cos (\phi-\phi_0),\\
\psi_2 &=\Omega_2 \cos \theta + a \sin \theta \cos (\phi-\phi_0),
\end{align}
\end{subequations}
with $\Omega_1=\sqrt{\frac{3}{4\pi}}\frac{2\Omega+\mu_c Z_1}{\beta_1+ \gamma_1^{\pm}\thermo{Z}_1+F}+\frac{F}{\beta_1+ \gamma_1^{\pm}\thermo{Z}_1+F} \Omega_2$. Hence, the flow is a superposition of a solid-body rotation and a dipole in each layer. Note that the dipole phases are necessarily the same.  The barotropic and baroclinic components of the solid-body rotation are given by, respectively,
\begin{align}
\Omega_t &=\frac 1 2 \sqrt{\frac{3}{4\pi}}\frac{2\Omega+\mu_c Z_1}{\beta_1+ \gamma_1^{\pm}\thermo{Z}_1+F}+\frac 1 2 \left\lbrack \frac{F}{\beta_1+ \gamma_1^{\pm}\thermo{Z}_1+F} +1\right\rbrack\Omega_2,\\
\Omega_c &=\frac 1 2 \sqrt{\frac{3}{4\pi}}\frac{2\Omega+\mu_c Z_1}{\beta_1+ \gamma_1^{\pm}\thermo{Z}_1+F}+\frac 1 2 \left\lbrack\frac{F}{\beta_1+ \gamma_1^{\pm}\thermo{Z}_1+F} -1 \right\rbrack \Omega_2.
\end{align}
Note that we have the simple relations $\Omega_t+\Omega_c=\Omega_1$, $\Omega_t-\Omega_c=\Omega_2$. Finally, the barotropic and baroclinic modes are given by
\begin{align}
\psi_t &= \Omega_t \cos \theta + \frac{F+(\beta_1+Z_1\gamma_1^{\pm})/2}{F+\beta_1+ \thermo{Z}_1\gamma_1^{\pm}} a \sin \theta \cos (\phi-\phi_0),\\
\psi_c &=\Omega_c \cos \theta - \frac{\beta_1+Z_1\gamma_1^{\pm}}{F+\beta_1+ \thermo{Z}_1\gamma_1^{\pm}} a \sin \theta \cos (\phi-\phi_0),
\end{align}

The angular momentum is given as usual by $\macro{L}=\beta_1\Omega_t/3$, and the energy by
\begin{equation}
\begin{split}
\macro{E}=  \frac{\beta_1}{6}\Omega_t^2+(\beta_1+2F)\frac{\Omega_c^2}{6} + \frac{\beta_1}{12} \Big\lbrack  \beta_1 \abs[\Big]{\frac{F_1}{\beta_1+ \thermo{Z}_1\gamma_1^{\pm}+F_1}}^2& \\
+ \beta_1  + F \abs[\Big]{\frac{F_1}{\beta_1+ \thermo{Z}_1\gamma_1^{\pm}+F_1}-1}^2\Big\rbrack a^2&.
\end{split}
\end{equation}
These relations show that $\Omega_2$ (and thus also $\Omega_1$) is directly fixed by the angular momentum $\macro{L}$, while the amplitude for the dipole $a$ is determined by the energy relation.

When $\thermo{Z}_1=\thermo{Z}_2=\thermo{Z}$, we recover the relation $\mu_c=-2\Omega/\thermo{Z}$, familiar from the one-layer case (see \cite{Herbert2012b}, where $\thermo{Z}$ is absorbed in the definition of $\mu$), which implies in particular that the first term in $\Omega_t,\Omega_c$ vanishes. In that case, for $\beta=\gamma_1^+$, the critical points of the variational problem lead to purely barotropic flows: $\psi_1=\psi_2=\psi_t$ and $\psi_c=0$. On the contrary, when $\beta=\gamma_1^-$, the critical points are purely baroclinic flows: $\psi_t=0, \psi_c \neq 0$.

\section{Thermodynamical Stability and Phase Diagrams}\label{stabsec}

So far we have only solved the equation defining the critical points of the mean-field variational problem in the energy-enstrophy limit. We now need to determine which of these critical points are actual maxima, minima or saddle-points of the mean-field equation.  This can be achieved by looking at the second variations of the grand-potential functional $\macro{J}[q_1,q_2]=\macro{S}[q_1,q_2]-\beta \macro{E}[q_1,q_2]-\mu\macro{L}[q_1,q_2]$. We have
\begin{widetext}
\begin{align}
\delta^2 \macro{J}[q_1,q_2] &= - \frac {1}{4}\int_{\domain} \frac{\delta q_1^2}{\thermo{Z}_1} d\vec{r} - \frac {1}{4}\int_{\domain} \frac{\delta q_2^2}{\thermo{Z}_2} d\vec{r}-\frac {\beta}{4} \int_{\domain} (\nabla \delta\psi_1)^2 d\vec{r} - \frac {\beta}{4} \int_{\domain} (\nabla \delta\psi_2)^2 d\vec{r}-\frac{\beta}{4} F \int_{\domain} (\delta\psi_1-\delta\psi_2)^2d\vec{r},\\
&= -\frac 1 4 \sum_{n=1}^{+\infty} \sum_{m=-n}^{n}\Big\lbrack  \frac{\abs{\delta q_{nm}^{(1)}}^2}{\thermo{Z}_1} + \frac{\abs{\delta q_{nm}^{(2)}}^2}{\thermo{Z}_2} +  \beta \beta_n \abs{\delta \psi_{nm}^{(1)}}^2 +\beta \beta_n \abs{\delta \psi_{nm}^{(2)}}^2\notag\\
&\phantom{-\frac 1 2 \sum_{n=1}^{+\infty} \sum_{m=-n}^{n}\Big\lbrack} +\beta F \abs{\delta \psi_{nm}^{(1)}}^2 +\beta F \abs{\delta \psi_{nm}^{(2)}}^2 -2\beta F \Re(\delta \psi_{nm}^{(1)} \delta {\psi_{nm}^{(2)}}^*) \Big\rbrack.
\end{align}
Clearly, for $\beta>0$, $\delta^2 \macro{J}[q_1,q_2]<0$ for any perturbation.

From \eqref{vortcoeff}, we have
\begin{subequations}
\begin{align}
\abs{\delta q_{nm}^{(1)}}^2 & = (\beta_n+F)^2 \abs{\delta\psi_{nm}^{(1)}}^2 + F^2 \abs{\delta\psi_{nm}^{(2)}}^2 -2F(\beta_n+F)\Re(\delta\psi_{nm}^{(1)}{\delta\psi_{nm}^{(2)}}^*),\\ 
\abs{\delta q_{nm}^{(2)}}^2 & = (\beta_n+F)^2 \abs{\delta\psi_{nm}^{(2)}}^2 + F^2 \abs{\delta\psi_{nm}^{(1)}}^2 -2F(\beta_n+F)\Re(\delta\psi_{nm}^{(1)}{\delta\psi_{nm}^{(2)}}^*). 
\end{align}
\end{subequations}
Therefore, we can write
\begin{equation}
\begin{split}
\delta^2 \macro{J}[q_1,q_2] = &-\frac 1 4 \sum_{n=1}^{+\infty} \sum_{m=-n}^{n} \Bigl\lbrack  a_n \abs{\delta \psi_{nm}^{(1)}}^2 + b_n \abs{\delta \psi_{nm}^{(2)}}^2 -2c_n \Re(\delta \psi_{nm}^{(1)} \delta {\psi_{nm}^{(2)}}^*) \Bigr\rbrack.
\end{split}
\end{equation}
\end{widetext}
with
\begin{align}
a_n &= (\beta_n+F)\beta + \frac{(\beta_n+F)^2}{\thermo{Z}_1} +\frac{F^2}{\thermo{Z}_2},\\
b_n &= (\beta_n+F)\beta +\frac{(\beta_n+F)^2}{\thermo{Z}_2} +\frac{F^2}{\thermo{Z}_1},\\
c_n &= F \left\lbrack \beta+ \frac{\beta_n+F}{\thermo{Z}_1}+\frac{\beta_n+F}{\thermo{Z}_2} \right\rbrack.
\end{align}

The quadratic form $\delta^2 \macro{J}[q_1,q_2]$ has a block diagonal form; to study its positive-definiteness, it remains to diagonalize the $2\times2$ blocks, which is achieved, for instance, by the change of variables:
\begin{align}
\delta \phi_{nm}^{(1)}&=\delta \psi_{nm}^{(1)}-\frac{c_n}{a_n}\delta \psi_{nm}^{(2)},\label{varchangeeq}\\
\delta \phi_{nm}^{(2)}&= \delta \psi_{nm}^{(2)}.
\end{align}
The quadratic form then reads
\begin{align}
\delta^2 \macro{J}[q_1,q_2] &= -\frac 1 4 \sum_{n=1}^{+\infty} \sum_{m=-n}^{n} \left(  a_n \abs{\delta \phi_{nm}^{(1)}}^2 + b_n' \abs{\delta \phi_{nm}^{(2)}}^2  \right).
\end{align}
with $b_n'=b_n-\frac{c_n^2}{a_n}$. Note that this change of variables breaks the symmetry $1 \leftrightarrow 2$.

\subsection{Grand-canonical stability}

The condition for grand-canonical stability is that the quadratic form $\delta^2 \macro{J}[q_1,q_2]$ be negative-definite for all perturbations $\delta q_1$, $\delta q_2$.
It is negative-definite if and only if, for all $n \in \N^*$, $a_n>0$ and $b_n'>0$. Lengthy but straightforward computations show that 
\begin{align}
b_n'&= \beta_n (\beta_n+2F)\frac{(\beta-\gamma_n^+)(\beta-\gamma_n^-)}{2 a_n}
\end{align}
so that $b_n'>0 \iff (\beta-\gamma_n^+)(\beta-\gamma_n^-)/a_n>0$, and the condition for grand-canonical stability can also be stated as $a_n>0$ and $(\beta-\gamma_n^+)(\beta-\gamma_n^-)>0$.

Let us start by assuming that $\beta > \gamma_1^+$. Then for $n \in \N^*$, $b_n'$ and $a_n$ have the same sign. Furthermore, direct computations show that $a_n(\gamma_n^+)>0$ (see appendix \ref{anplusappendix}). As each $a_n(\beta)$ is an increasing function, we have $a_n(\beta)>0$, and thus also $b_n'(\beta)>0$, for all $n \in \N^*$. Thus, for $\beta > \gamma_1^+$, $\delta^2 \macro{J}[q_1,q_2] <0$ for every perturbation $(\delta q_1,\delta q_2)$. This is the condition of \emph{grand-canonical stability}, which implies in particular \emph{microcanonical stability} \cite{Chavanis2009}.

Note that when $\beta = \gamma_1^+$, all the above remains true, except that now $b_1'(\beta)=0$. The quadratic form is negative but not definite anymore. In fact, it is degenerate; its radical is given by all perturbations $\delta \psi_2 \in \Ker(\Delta+\beta_1 \Id)$. This possibility has been studied in detail in the case of barotropic flows~\citep{Herbert2012a}. Physically, it means that in the grand-canonical ensemble, we have a metastable state, and spontaneous transitions are possible in the upper layer between zonal flows and dipoles. This is accompanied by a similar transition in the lower layer, as can be seen by the form of the change of variables \eqref{varchangeeq}.

It remains to treat the case $\beta < \gamma_1^+$. We either have $\beta \leq \gamma_1^-$ or $\gamma_1^- < \beta < \gamma_1^+$. We know that if $\gamma_n^-<\beta<\gamma_n^+$ for some $n\geq 1$, then $a_n b_n'<0$ and we have grand-canonical instability. 
Besides, it can be proved that $a_n(\gamma_n^-)<0$ (see appendix \ref{anminusappendix}). In particular, for all $\beta \leq \gamma_1^-$, we have grand-canonical instability because $a_1(\beta)<0$.

As a conclusion, the quadratic form $\delta^2 \macro{J}$ is negative definite for $\beta>\gamma_1^+$ (grand-canonical stability, and therefore also microcanonical stability), not sign-definite when $\beta<\gamma_1^+$ (grand-canonical instability), and degenerate and negative when $\beta=\gamma_1^+$ (grand-canonical metastability).

\subsection{Microcanonical stability}

The microcanonical stability analysis is a little more subtle, because we must restrict ourselves to the perturbations which conserve energy and angular momentum at first order. The first order variations of these quantities is given by
\begin{align}
\delta \macro{E} &= \frac 1 2 \int_{\domain} \delta q_1 \psi_1 + \frac 1 2 \int_{\domain} \delta q_2 \psi_2,\\
\delta \macro{L} &= \frac 1 2 \int_{\domain} \delta q_1 \cos \theta + \frac 1 2 \int_{\domain} \delta q_2 \cos \theta.
\end{align}

The condition for microcanonical stability is that the quadratic form $\delta^2 \macro{J}[q_1,q_2]$ be negative-definite for all perturbations $\delta q_1$, $\delta q_2$ which preserve the energy and angular momentum constraints at first order. Clearly, grand-canonical stability implies microcanonical stability, so that we already know that the solutions for $\beta > \gamma_1^+$ are stable in the microcanonical ensemble. It remains to check if solutions which are not stable in the grand-canonical ensemble ($\beta<\gamma_1^+$) may become stable in the microcanonical ensemble.

Let us first assume that $\beta \notin \pseudospectrum$; the background flow is a solid-body rotation, so that it suffices that the perturbations be orthogonal to the solid-body rotations for them to preserve the invariants at first order. When $\beta < \gamma_1^-$, we have $a_1(\beta)<0$ and a dipole perturbation $\delta \phi_{1}$ will destabilize the flow ($\delta^2 \macro{J} >0$), while $\delta \macro{E}=0, \delta \macro{L}=0$. When $\gamma_1^- < \beta < \gamma_1^+$, we have $a_1(\beta)b_1'(\beta)<0$, and a dipole perturbation in the appropriate variable will also destabilize the flow while conserving energy and angular momentum at first order.
Finally, all the solid-body rotations with $\beta \in ]-\infty,\gamma_1^+[ \setminus \pseudospectrum$ are saddle points of the constrained variational problem.

For $\beta=\gamma_n^{\pm}$, with $n>1$, the same reasoning as above applies: as the background flow is the sum of a solid-body rotation and a Laplacian eigenvector of order $n$, dipole perturbations are orthogonal to the background flow and thus preserve the constraints at first order: $\delta \macro{E}=0, \delta \macro{L}=0$. Hence, when $\gamma_n^{\pm} < \gamma_1^-$ (this is always the case for $\gamma_n^-$), $a_1(\gamma_n^{\pm})<0$ and it suffices to consider a dipole perturbation for $\delta \phi_1$. When $\gamma_1^- < \gamma_n^{+} < \gamma_1^+$ (which happens if there is root intertwining), $a_1(\gamma_n^+)b_1'(\gamma_n^{+})<0$ and one needs to consider a dipole perturbation either in $\delta \phi_1$ or $\delta \phi_2$. In all cases, the critical points are also saddle points of the microcanonical variational problem.

It remains to treat the cases $\beta=\gamma_1^{\pm}$. In both cases, $b_1'(\beta)=0$ (and for $n>1$, $b_n'(\gamma_1^+)>0$ while the sign of $b_n'(\gamma_1^-)$ depends on whether there is root intertwining or not --- if not, they are all positive). When $\beta=\gamma_1^-$, $a_1(\gamma_1^-)<0$ and one may build a dipole perturbation $\delta \phi_1$ while $\delta \phi_2=0$ which conserves the energy and angular momentum at first order. Hence in that case, the critical points are saddle-points of the microcanonical variational problem.

For $\beta=\gamma_1^+$, the quadratic form $\delta^2 \macro{J} $ is negative but not definite: $b_1'(\beta)=0$. It is degenerate in the sense that perturbing the flow with an energy preserving dipole $\delta \phi_2$ while $\delta\phi_1=0$ does not modify the grand potential up to second order. The dipoles are therefore metastable states.

It may be useful to give more explanations as to why energy preserving dipole perturbations destabilize the background flow --- which is itself the sum of a dipole and a solid-body rotation --- in the case $\beta=\gamma_1^-$, but not in the case $\beta=\gamma_1^+$. Note that in both cases, the critical points of the variational problem correspond to dipoles in the two layers, with the same phase and a proportionality relation between the amplitudes. There are energy-preserving dipole perturbations which conserve these constraints, and dipole perturbations which break them, either because they introduce a phase difference between the two layers or because they modify the ratio of the amplitudes (or both). The first kind of perturbations does not modify the grand potential, both for $\beta=\gamma_1^-$ and $\beta=\gamma_1^+$. Hence, in both cases, there is a whole family of critical points with the same energy, the same angular momentum, and the same entropy. This is a one parameter family described by the phase of the dipole, common to the two layers. The second kind of perturbations, on the contrary, leads to dipole states which, even if they have the same amplitude (necessary to preserve the energy), may lose the phase constraint. These states can have higher or lower entropy than the critical points. When $\beta=\gamma_1^-$, they have a higher entropy and the critical points are saddle points of the microcanonical variational problem, and when $\beta=\gamma_1^+$, they have a lower entropy and the critical points are maxima of the microcanonical variational problem.

Finally, the stability properties are the same in the microcanonical ensemble and the grand-canonical ensemble, except for the metastability properties in the case $\beta=\gamma_1^+$. This is a case of marginal ensemble equivalence, similar to the case of a one-layer model~\cite{Herbert2012a}.

\subsection{Stability and root intertwining}

In the above analysis, we have proved (grand-canonical and microcanonical) instability for $\beta<\gamma_1^+$ using essentially dipole perturbations, because either $a_1(\beta)<0$ or $b_1'(\beta)<0$. But the dipole perturbations are large-scale, belong to a low-dimensional subspace of phase space and therefore might not be generated spontaneously in practice. In most cases, however, there are other unstable directions and other types of destabilizing perturbations. 

In general, if $\gamma_{n+1}^- < \beta < \gamma_n^-$, $a_p(\beta)<0$ for $1 \leq p \leq n$ and any perturbation in $\bigoplus_{p=1}^n \Ker (\Delta + \beta_p \Id)$ will destabilize the flow, in the grand canonical ensemble. In the microcanonical ensemble, this remains true when $\beta \notin \pseudospectrum$, but only those perturbations which conserve energy and angular momentum at first order remain when $\beta \in \pseudospectrum$. Thus, roughly speaking, the smaller $\beta$ is, the less stable the critical point is. This source of instability exists in the barotropic case~\citep{Herbert2012a} and does not depend on whether there is root intertwining or not.

There is a second source of instability which acts when $\gamma_n^- < \beta < \gamma_n^+$ and $a_n(\beta)>0$. This source of instability becomes more important as there is more root intertwining. For instance, the dipole critical points obtained for $\beta=\gamma_1^-$ can only be destabilized by dipoles when there is no root intertwining, but it is not so when there is. In cases 2 (when $\thermo{Z}_1=\thermo{Z}_2$) and 4 ($\thermo{Z}_1 \neq \thermo{Z}_2$), we have $\gamma_1^- < \gamma_p^+ < \cdots < \gamma_1^+$, and perturbations in $\bigoplus_{k=1}^p \Ker (\Delta + \beta_k \Id)$ will destabilize the flow (taking into account the energy and angular momentum conservation if we are working in the microcanonical ensemble). In the case $\thermo{Z}_1 \neq \thermo{Z}_2$, root intertwining occurs at least for $n$ large enough. This means that for $\beta$ small enough, the critical points can be destabilized by arbitrary small scale perturbations.

\subsection{Phase Diagrams}

\begin{figure}
\includegraphics[width=0.45\linewidth]{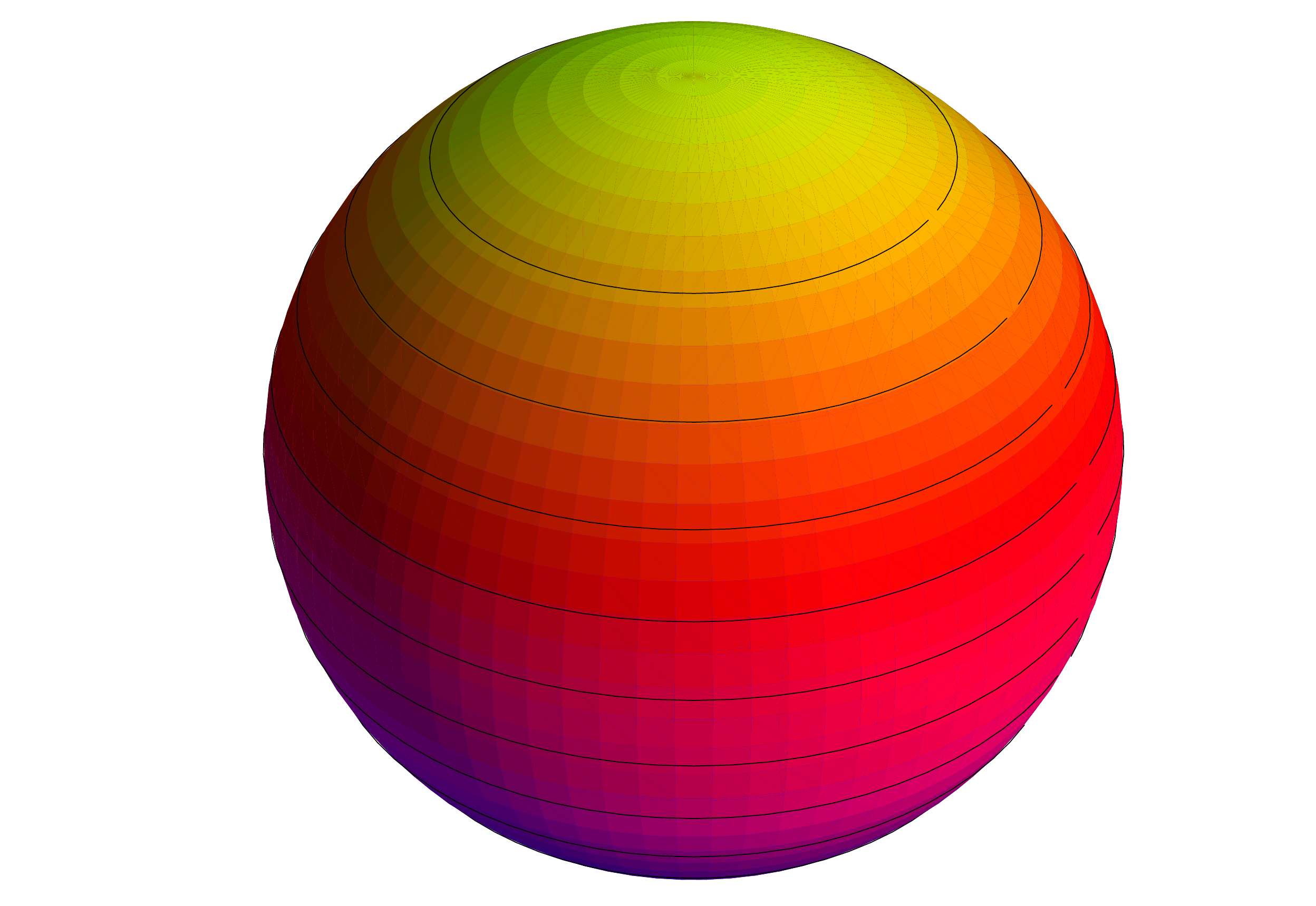}
\includegraphics[width=0.45\linewidth]{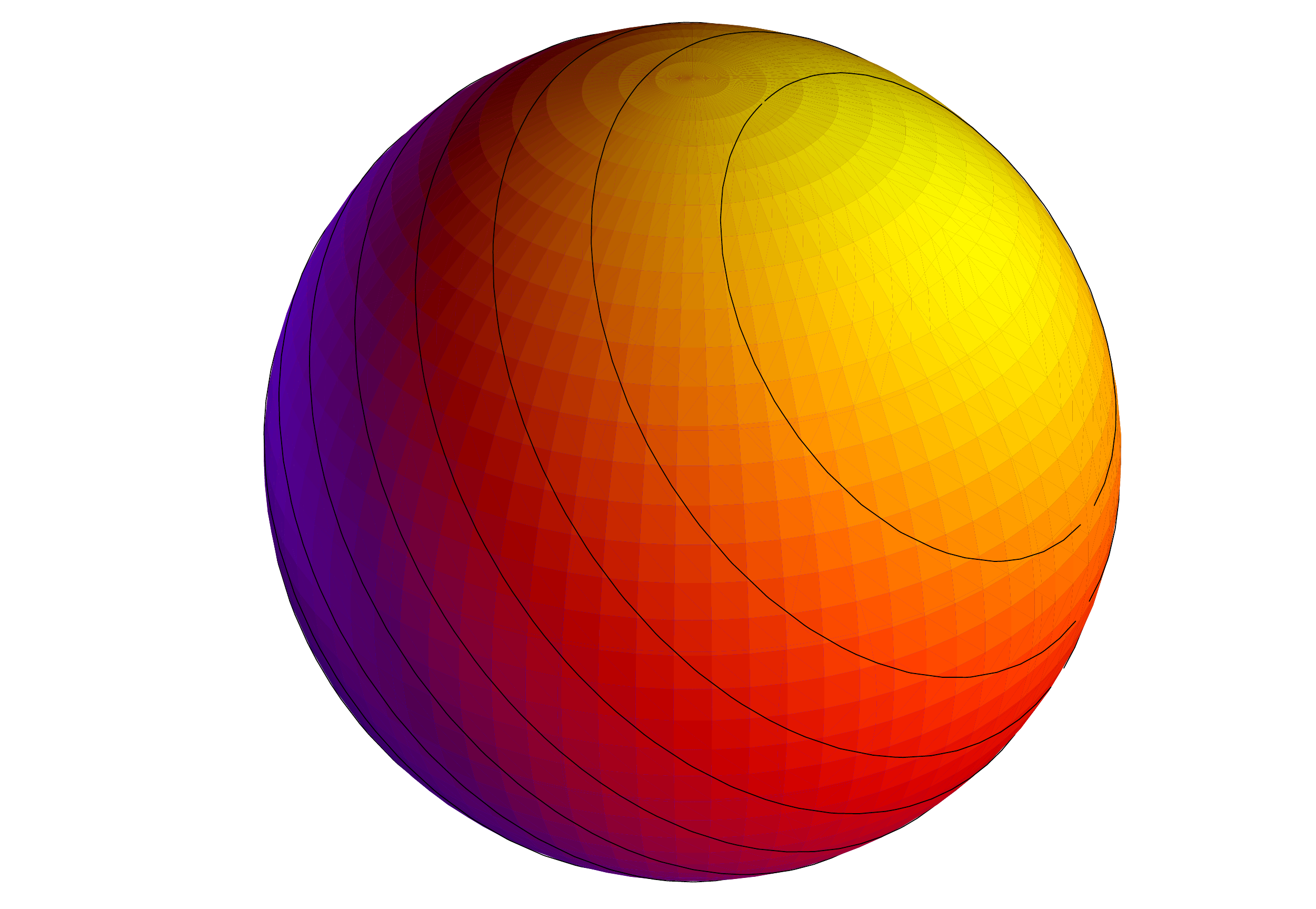}
\caption{(Color online) Horizontal structure of the flow: vorticity field at statistical equilibrium for one layer. There are two possible flow configurations at statistical equilibrium: solid-body rotation (left) or dipole (right).}\label{horeqfig}
\end{figure}
\begin{figure}[tbhp]
\includegraphics[width=\linewidth]{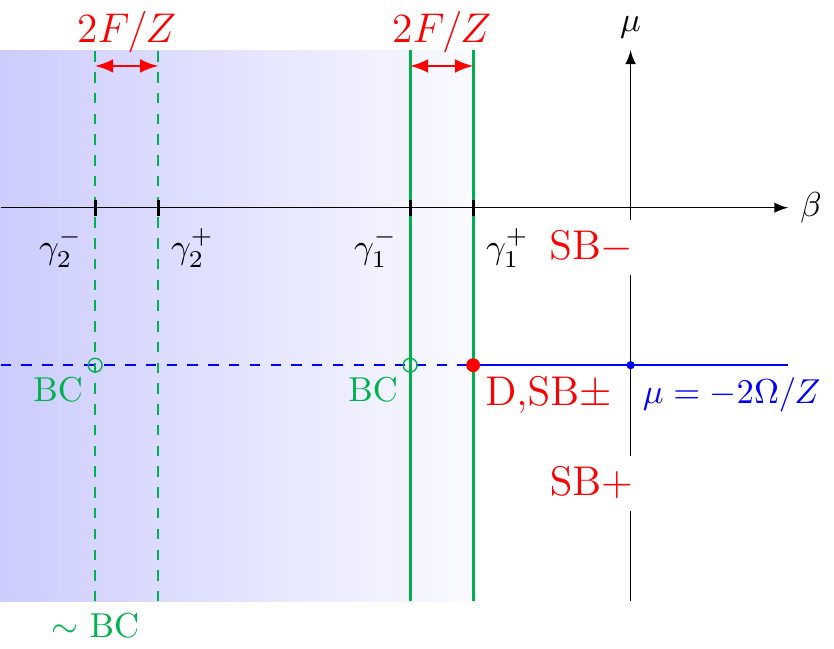}
\caption{(Color online) Phase diagram for two-layer quasi-geostrophic flows on a sphere in the grand-canonical ensemble, in the case $Z_1=Z_2=Z$. For simplicity, we are drawing the phase-diagram in the case where the roots are not intertwined ($F < 2$). The shaded area corresponds to unstable states.}\label{gcphasediaghom}
\end{figure}
\begin{figure}
\includegraphics[width=\linewidth]{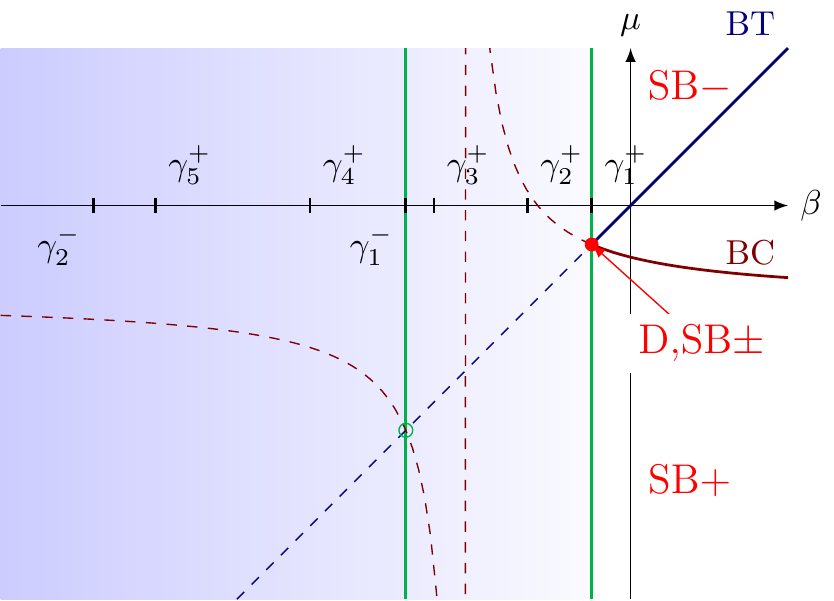}
\caption{(Color online) Phase diagram for two-layer quasi-geostrophic flows on a sphere in the grand-canonical ensemble, in the case $Z_1 \neq Z_2$. The shaded area corresponds to unstable states. BT (resp. BC) indicates the loci of purely barotropic (resp. baroclinic) flows.}\label{gcphasediagshear}
\end{figure}

All the above results can be summarized in the form of the \emph{phase-diagram} of the system. In the grand-canonical ensemble, the relevant variables are the Lagrange parameters $\beta$ and $\mu$. For each value of these parameters, we have critical points as described in section \ref{meanfieldeqsolsec}. We have seen above that these critical points correspond to actual maxima of the grand potential only when $\beta \geq \gamma_1^+$; hence the only equilibrium states are solid-body rotations (in each layer) or dipoles (with the same phase in each layer). The horizontal structure of the flow for the equilibrium states is depicted on Fig. \ref{horeqfig}.
These states are denoted on Figs. \ref{gcphasediaghom} and \ref{gcphasediagshear} by SB+ and SB- for co-rotating and counter-rotating solid-body rotations, respectively, and D for dipoles. All the critical points in the shaded area (left of the $\beta=\gamma_1^+$ vertical line) correspond to saddle points of the grand-canonical variational problem.
The phase diagram is drawn in Fig. \ref{gcphasediaghom} for the case of a vertically homogeneous fine-grained enstrophy profile: $Z_1=Z_2$. In this case, all the points in the phase diagram except those on the vertical lines $\beta=\gamma_n^-$ are pure barotropic flows. On the contrary, the (unstable) states on the vertical lines $\beta=\gamma_n^-$ correspond to quasi-baroclinic flows: they are the superposition of a barotropic solid-body rotation and a baroclinic flow. The barotropic solid-body rotation component vanishes at the intersection with the line $\mu=\mu^*(\beta)=-2\Omega/Z$, and there we have pure baroclinic flows. Note that on the vertical (solid green) lines $\beta=\gamma_1^\pm$, solutions of the mean field equation only exist at the intersection with the line $\mu=\mu^*(\beta)=-2\Omega/Z$.
When $\thermo{Z}_1 \neq \thermo{Z}_2$, the phase diagram (see Fig. \ref{gcphasediagshear}) has a qualitatively similar structure. In particular, the equilibrium states remain solid body rotations and dipoles. The main difference is that an arbitrary point in the diagram now corresponds to both non-vanishing barotropic and baroclinic components. The purely barotropic flows are located on the blue straight line, while the purely baroclinic flows correspond to the red curve. A notable difference is that pure barotropic and pure baroclinic flows are now attainable at statistical equilibrium. As $\thermo{Z}_1-\thermo{Z}_2 \to 0$, the lines of pure barotropic and pure baroclinic flows move closer, until they collapse for $\thermo{Z}_1=\thermo{Z}_2$ and form the line $\mu=-2\Omega/Z$, where the flow vanishes except at the intersections with lines $\beta=\gamma_n^-$ or $\beta=\gamma_1^+$.

\section{Discussion}\label{discsec}

In the previous sections, we have obtained analytically a class of equilibrium states for two-layer quasi-geostrophic flows characterized by a linear relation between vorticity and stream function. In addition, we have also obtained all the critical points of the corresponding variational problem, including saddle-points of the variational problem. We have also constructed the phase diagrams of the system. Note that we restricted ourselves here to the case of a linear vorticity-stream function relation. For 2D flows on the sphere, Qi and Marston~\cite{Qi2014} have performed a perturbative computation enforcing weakly the constraint fo conservation of the quartic Casimir invariant, and obtained that the presence of a weak non-linearity in the vorticity-stream function leads to a sharpening or to a weakening of the vorticity profile in the vortex cores. We can expect this result to carry over to the two-layer quasi-geostrophic flows considered here; therefore the repartition of energy between the barotropic and the baroclinic modes at statistical equilibrium should not be too much affected by the presence of a weak non-linearity.

These results indicate that the behavior of the system on the horizontal scales is similar to that of the one-layer (barotropic) system. The phase space is made of two copies of the one-layer phase space (formally, it is the cartesian product $L^2(\domain) \times L^2(\domain)$). It is convenient to decompose the phase space in terms of the eigenspaces of the Laplace-Beltrami operator: $L^2(\domain)=\bigoplus_{k=1}^{+\infty} V_k$, with $V_k = \Ker(\Delta+\beta_k\Id)$. In the one-layer case, the equilibrium states corresponding to a linear vorticity-stream function relation all lie in $V_1$; in other words, the \enquote{final result} of the inverse cascade is expected to be a complete condensation of the energy in the gravest modes. In the two-layer case considered here, all the energy is analogously expected to condense in $V_1 \times V_1$. Hence, the stratification does not really modify the nonlinear transfers of energy in each layer on the horizontal.

The nonlinear transfers of energy on the vertical direction --- which of course is not accounted for at all in a barotropic model --- are more complicated. The standard, phenomenological theory for two-layer quasi-geostrophic turbulence~\cite{Salmon1978,Rhines1979,Salmon1980} (see also \cite[chap. 9]{VallisBook}) is based on the fact that, at large scales (larger than the radius of deformation), the dynamical equation for the barotropic mode resembles that of 2D turbulence, while the dynamics of the baroclinic mode resembles a passive scalar. Hence, one expects a direct cascade of baroclinic energy and an inverse cascade of barotropic energy. As the forcing in the oceans and in the atmosphere is essentially baroclinic, the traditional picture which emerges is that energy is injected mostly at large scales in the baroclinic mode, cascades downscale until reaching the radius of deformation, where it is in major part transferred to the barotropic mode and cascades upscale until being dissipated by boundary layer friction. Note that at large scales, the inverse cascade of barotropic energy can be arrested by the effect of Rossby waves, which leads to preferential formation of zonal flows~\cite{Rhines1975,Vallis1993}. The energy which is not transferred to the barotropic mode at the deformation scale, and not cascaded upscale in the barotropic mode, is scattered downscale until it reaches the Ozmidov scale, where isotropy recovers and turbulence becomes 3D. 
Statistical mechanics sheds some light on the nonlinear processes which transfer energy from the baroclinic mode to the barotropic mode. 
Previous studies, based either on the Galerkin truncated approach for two-layer flows~\cite{Salmon1976} or the MRS approach for continuous stratification~\cite{Venaille2012a} have stressed that there is a natural tendency for \emph{barotropization} of the flow, and that statistical equilibria should be barotropic at large scales. 
Our results complete this picture by confirming the general tendency towards barotropization and allowing a precise discussion based on the phase diagrams. In the context of the two-layer model, it is clear that the conservation of fine-grained enstrophy in each layer will put strong constraints on the extent to which barotropization can be achieved. Indeed, purely barotropic flows necessarily have $\thermo{Z}_1=\thermo{Z}_2$ (because $\psi_1=\psi_2$). 
Nevertheless, enstrophy is expected to be transferred downscale and even if $\thermo{Z}_1 \neq \thermo{Z}_2$, it may happen that most of the difference is in fact in the small scales, while the amount of enstrophy in the large scales is roughly the same in the two-layers, thereby allowing for complete barotropization in the large scales. The mean-field theory used here is precisely concerned not with the fine-scale fields but with coarse-grained ones. Therefore, at a coarse-grained level describing the large scales, it may be that $\overline{\psi}_1=\overline{\psi}_2$. Our results state that when $\thermo{Z}_1=\thermo{Z}_2$, the barotropization process reaches its full extent; i.e. all the statistical equilibria are purely barotropic and no energy remains in the baroclinic mode. When $\thermo{Z}_1\neq \thermo{Z}_2$, purely barotropic flows are still attainable at a coarse-grained level due to the mechanism explained above, but purely baroclinic flows are also possible, and in general the statistical equilibria still contain some baroclinic energy. The fraction of energy blocked in the baroclinic mode is proportional to $(\thermo{Z}_1-\thermo{Z}_2)^2$ -- but also depends on $\beta$ and $\mu$ of course.

As noted in~\cite{Venaille2012a}, rotation tends to favor barotropization. This can be expected based on the remark that as planetary vorticity $f$ increases, it becomes the dominant term in total vorticity $q_i$. Therefore rotation tends to homogenize the vertical distribution of fine-grained enstrophy as well, and increasing $f$ decreases the fraction of baroclinic energy at statistical equilibrium. This can be seen on the phase diagram (Fig. \ref{gcphasediagshear}), where increasing $f$ brings the curve of pure baroclinic flows towards the $\mu=-\infty$ region. As mentioned in~\cite{Venaille2012a}, this effect is probably counter-balanced after a certain threshold by the effect of waves: as $f$ increases, the propagation time of Rossby waves become smaller, until it becomes comparable to the eddy turnover time at the same scale. Then, we enter a regime dominated by wave dynamics, where the dynamics may not be mixing enough to ensure that the predictions of statistical mechanics will be observed. As a matter of fact, it has been observed in numerical simulations~\cite{Venaille2012a} that Rossby waves break the tendency towards barotropization.

As a summary, statistical mechanics predicts that the barotropization process is maximal when starting from a vertically homogeneous (or nearly so) fine-grained enstrophy profile, while it is only partial, and can range from zero to maximal, when starting from a vertically sheared fine-grained situation. These results could be of relevance to better understand the role of balanced modes in vertical mixing in rotating/stratified fluids like the ocean, as opposed to, for instance, mixing induced by internal gravity waves.

\appendix

\section{Sign of the eigenvalues $a_n(\gamma_n^\pm)$ for grand potential second-variation $\delta^2 \macro{J}$}

In this appendix, we give the elementary proofs that $\forall F \in \R_+, \forall Z_1,Z_2 \in \R_+^*, a_n(\gamma_n^+)>0$ and $a_n(\gamma_n^-)<0$.

\subsection{Proof that $a_n(\gamma_n^+)>0$}\label{anplusappendix}

Direct computations show that
\begin{align}
\begin{split}
a_n(\gamma_n^+) &= \frac{\beta_n+F}{2 Z_1 Z_2} \sqrt{(\beta_n+F)^2(Z_1-Z_2)^2+4F^2 Z_1 Z_2}\\
&\phantom{=}+\frac{(\beta_n+F)^2}{2Z_1}-\frac{(\beta_n+F)^2}{2Z_2}+\frac{F^2}{Z_2},
\end{split}\\
&> \frac{(\beta_n+F)^2}{2} \underbrace{\Big\lbrack \bigabs{\frac{1}{Z_2}-\frac{1}{Z_1}} + \Big( \frac{1}{Z_1} - \frac{1}{Z_2}\Big)\Big\rbrack}_{\geq 0} +\frac{F^2}{Z_2}\\
&>0.
\end{align}

\subsection{Proof that $a_n(\gamma_n^-)<0$}\label{anminusappendix}

Here the proof is still elementary but slightly longer. Let us start by fixing arbitrary $F \geq 0$ and $Z_2>0$. Note that Taylor expansions give the limits of $a_n(\gamma_n^-)$, seen as a function of $Z_1$, at $0$ and $+\infty$:
\begin{align}
\lim_{Z_1 \to 0} a_n(\gamma_n^-) &= 0,\\
\lim_{Z_1 \to +\infty} a_n(\gamma_n^-) &= -\frac{\beta_n(\beta_n+2F)}{Z_2} <0.
\end{align}

Direct computation shows that
\begin{align}
\frac{\partial a_n(\gamma_n^-)}{\partial Z_1} &= \frac{\beta_n+F}{\sqrt{(\beta_n+F)^2(Z_1-Z_2)^2+4F^2 Z_1 Z_2}} \frac{Z_2}{Z_1} a_n(\gamma_n^-).
\end{align}
In other words, $a_n(\gamma_n^-)$, as a function of $Z_1$, satisfies a linear differential equation of order one. As it is not the null function, it does not vanish and has a constant sign; either $\forall Z_1 \in \R_+^*, a_n(\gamma_n^-)>0$ or $\forall Z_1 \in \R_+^*, a_n(\gamma_n^-)<0$. Let us assume that it is always positive. Then the logarithm $\ln a_n(\gamma_n^-)$ is well defined and is an increasing function of $Z_1$; hence so is $a_n(\gamma_n^-)$ itself. But its limit at $+\infty$ is negative, which is a contradiction. Hence $\forall Z_1 \in \R_+^*, a_n(\gamma_n^-)<0$.

\section{Phase diagram for the equivalent barotropic equation}

A case simpler than the full two-layer quasi-geostrophic equations is the so-called \emph{equivalent barotropic equation}:
\begin{align}
\partial_t q + \{\psi, q\} &= 0, \qquad q=-\Delta \psi+\frac{\psi}{R^2}+f.
\end{align}
This case was treated in details in~\cite{Herbert2012b} in the presence of a bottom topography. Let us take the opportunity to correct a typo in~\cite{Herbert2012b}: Eq.~52 should be
\begin{equation}
L = \langle (q-f) \cos \theta\rangle = \frac{1}{\sqrt{12\pi}} \scalprod{(q-f)}{Y_{10}}.
\end{equation}

We recall here the effect of the Rossby deformation radius $R$ on the phase diagram without the complication of the bottom topography.

\subsection{Equilibrium states}

Let $\lambda=\beta+\frac{1}{R^2}$ and $\lambda_n=\beta_n+\frac{1}{R^2}$.
The equilibrium states are as follows.

When $\lambda \notin \Sp \Delta$, we have a solid-body rotation $\psi=\Omega_* \cos \theta$ with $\Omega_*=-\frac{2\Omega+\mu}{\beta+\lambda_1}$. The energy, angular momentum and generalized entropy are given by
\begin{align}
E &= \frac{\lambda_1}{6} \Omega_*^2 = \frac{\lambda_1}{6} \left(\frac{2\Omega+\mu}{\beta+\lambda_1}\right)^2,\\
L &= \frac{\lambda_1}{3} \Omega_*     = -\frac{\lambda_1}{3} \frac{2\Omega+\mu}{\beta+\lambda_1},\\
S &= -\frac{1}{6} \left(\frac{2\Omega\beta-\mu\lambda_1}{\beta+\lambda_1}\right)^2.
\end{align}

When $\lambda=-\beta_1$, we only have a solution if $\mu=-2\Omega$. Then $\psi=\Omega_*\cos \theta+\gamma_c \sin\theta\cos \phi + \gamma_s\sin\theta\sin\phi$. The energy, angular momentum and generalized entropy are given by
\begin{align}
E &= \frac{\lambda_1}{6} \lbrack \Omega_*^2+\gamma_c^2+\gamma_s^2\rbrack,\\
L &= \frac{\lambda_1}{3} \Omega_*,\\
S &= -\frac{1}{6} \lbrack (\lambda_1\Omega_*+2\Omega)^2+\lambda_1^2(\gamma_c^2+\gamma_s^2) \rbrack.
\end{align}
Introducing the energy of a solid-body rotation with angular momentum $L$, $E_*(L)=3L^2/(2\lambda_1)$, and the phase $\phi_0$ such that $\gamma_c=\sqrt{6(E-E_*(L))/\lambda_1}\cos \phi_0, \gamma_s=\sqrt{6(E-E_*(L))/\lambda_1}\sin \phi_0$, we can write the equilibrium state as in \cite{Herbert2012a,Herbert2012b}
\begin{equation}
\psi = \frac{3L}{\lambda_1} \left\lbrack \cos\theta + \sqrt{\frac{E}{E_*(L)}-1} \sin \theta \cos (\phi-\phi_0)\right\rbrack.
\end{equation}

When $\lambda=-\beta_n, n>1$, we have other critical points but they are saddle points of the generalized entropy functional \cite{Herbert2012b}. On the contrary, the two critical points discussed above are maxima of the generalized entropy functional.

\subsection{Phase Diagrams and Thermodynamic Properties}

\begin{figure}[tbhp]
\includegraphics[width=\linewidth]{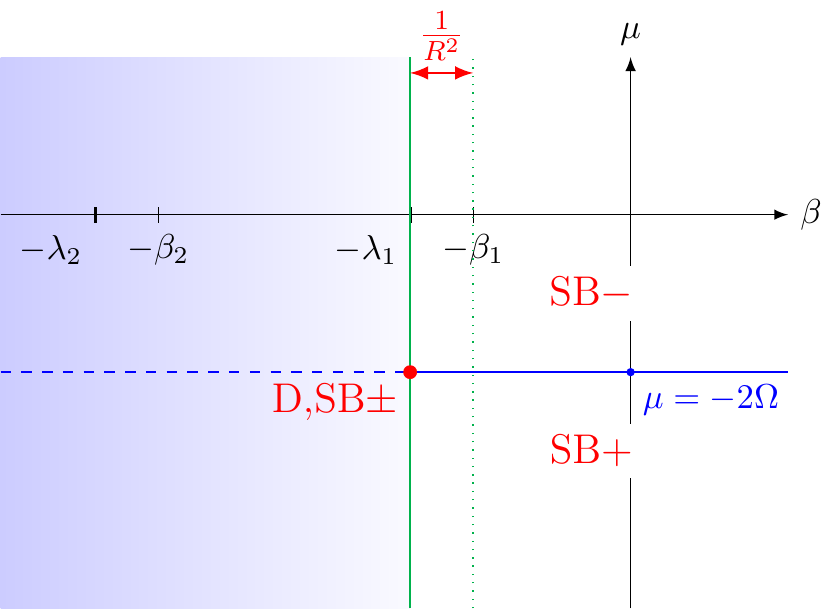}
\caption{(Color online) Grand-canonical phase diagram for the equivalent barotropic equation. All the flows on the left-hand side of the solid green line are unstable (shaded area). At the critical point (red dot), we have both dipoles and solid-body rotations. Above the solid blue line, we have counter-rotating solid-body rotations and below the line, co-rotating solid-body rotations. The effect of the Rossby deformation radius is to move the green line from its former position for $R=\infty$ (dotted green line) towards negative $\beta$. }\label{gcphasediagfig}
\end{figure}

\begin{figure}[tbhp]
\includegraphics[width=\linewidth]{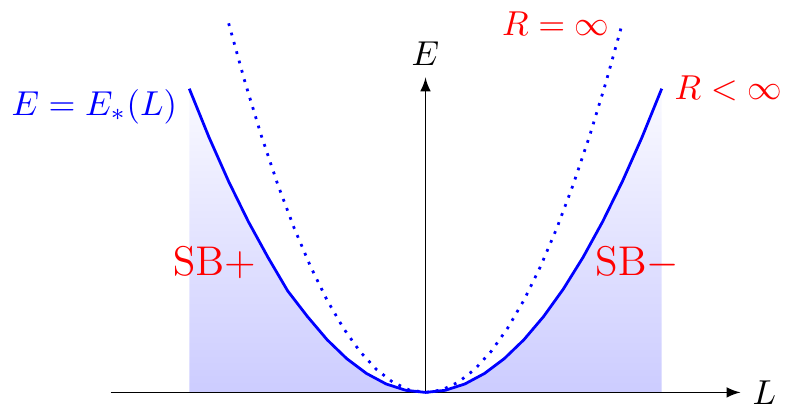}
\caption{(Color online) Microcanonical phase diagram for the equivalent barotropic equation. On the parabola $E=E_*(L)$, we have the solid-body rotations, while the dipole flows are above the parabola. The parabola corresponding to an infinite Rossby deformation radius is drawn with a dotted line. The effect of the Rossby deformation radius is thus simply to distort the parabola. The shaded area corresponds to inaccessible states.}\label{mcphasediagfig}
\end{figure}

The solutions found in the previous paragraph can be organized in a phase diagram. Figure \ref{gcphasediagfig} shows the phase diagram in the grand-canonical ensemble (compare to Fig.~2 of~\cite{Herbert2012a}) while Fig.~\ref{mcphasediagfig} shows the phase diagram in the microcanonical ensemble (compare to Fig.~3 of~\cite{Herbert2012a}).

One may also compute the thermodynamic potentials: we find
\begin{align}
\thermo{S}(\thermo{E},\thermo{L}) &= -\lambda_1 \thermo{E} -2\Omega \thermo{L} -\frac 2 3 \Omega^2,\\
\thermo{J}(\beta,\mu) &= -\frac{\lambda_1}{6}(\beta+\lambda_1)\Omega_*^2 -\frac{\lambda_1}{3}(2\Omega+\mu)\Omega_*-\frac{2}{3}\Omega^2,\\
&= \frac{\lambda_1}{6} \frac{(2\Omega+\mu)^2}{\beta+\lambda_1}-\frac{2}{3}\Omega^2.
\end{align}

In conclusion, a finite Rossby deformation radius just shifts or deforms the phase diagrams, but it does not change the overall structure, nor does it have an impact on the thermodynamic properties of the system.


\begin{thebibliography}{55}
\expandafter\ifx\csname natexlab\endcsname\relax\def\natexlab#1{#1}\fi
\expandafter\ifx\csname bibnamefont\endcsname\relax
  \def\bibnamefont#1{#1}\fi
\expandafter\ifx\csname bibfnamefont\endcsname\relax
  \def\bibfnamefont#1{#1}\fi
\expandafter\ifx\csname citenamefont\endcsname\relax
  \def\citenamefont#1{#1}\fi
\expandafter\ifx\csname url\endcsname\relax
  \def\url#1{\texttt{#1}}\fi
\expandafter\ifx\csname urlprefix\endcsname\relax\def\urlprefix{URL }\fi
\providecommand{\bibinfo}[2]{#2}
\providecommand{\eprint}[2][]{\url{#2}}

\bibitem[{\citenamefont{Monin and Yaglom}(1979)}]{MoninBook}
\bibinfo{author}{\bibfnamefont{A.~S.} \bibnamefont{Monin}} \bibnamefont{and}
  \bibinfo{author}{\bibfnamefont{A.~M.} \bibnamefont{Yaglom}},
  \emph{\bibinfo{title}{{Statistical Fluid Mechanics}}}
  (\bibinfo{publisher}{MIT Press}, \bibinfo{year}{1979}).

\bibitem[{\citenamefont{McWilliams}(1984)}]{McWilliams1984}
\bibinfo{author}{\bibfnamefont{J.~C.} \bibnamefont{McWilliams}},
  \bibinfo{journal}{J. Fluid Mech.} \textbf{\bibinfo{volume}{146}},
  \bibinfo{pages}{21} (\bibinfo{year}{1984}).

\bibitem[{\citenamefont{Marteau et~al.}(1995)\citenamefont{Marteau, Cardoso,
  and Tabeling}}]{Marteau1995}
\bibinfo{author}{\bibfnamefont{D.}~\bibnamefont{Marteau}},
  \bibinfo{author}{\bibfnamefont{O.}~\bibnamefont{Cardoso}}, \bibnamefont{and}
  \bibinfo{author}{\bibfnamefont{P.}~\bibnamefont{Tabeling}},
  \bibinfo{journal}{Phys. Rev. E} \textbf{\bibinfo{volume}{51}},
  \bibinfo{pages}{5124} (\bibinfo{year}{1995}).

\bibitem[{\citenamefont{Mininni and Pouquet}(2010)}]{Mininni2010b}
\bibinfo{author}{\bibfnamefont{P.~D.} \bibnamefont{Mininni}} \bibnamefont{and}
  \bibinfo{author}{\bibfnamefont{A.}~\bibnamefont{Pouquet}},
  \bibinfo{journal}{Phys. Fluids} \textbf{\bibinfo{volume}{22}},
  \bibinfo{pages}{035106} (\bibinfo{year}{2010}).

\bibitem[{\citenamefont{Mininni et~al.}(2011)\citenamefont{Mininni, Dmitruk,
  Matthaeus, and Pouquet}}]{Mininni2011a}
\bibinfo{author}{\bibfnamefont{P.~D.} \bibnamefont{Mininni}},
  \bibinfo{author}{\bibfnamefont{P.}~\bibnamefont{Dmitruk}},
  \bibinfo{author}{\bibfnamefont{W.~H.} \bibnamefont{Matthaeus}},
  \bibnamefont{and} \bibinfo{author}{\bibfnamefont{A.}~\bibnamefont{Pouquet}},
  \bibinfo{journal}{Phys. Rev. E} \textbf{\bibinfo{volume}{83}},
  \bibinfo{pages}{016309} (\bibinfo{year}{2011}).

\bibitem[{\citenamefont{Kraichnan}(1967)}]{Kraichnan1967}
\bibinfo{author}{\bibfnamefont{R.~H.} \bibnamefont{Kraichnan}},
  \bibinfo{journal}{Phys. Fluids} \textbf{\bibinfo{volume}{10}},
  \bibinfo{pages}{1417} (\bibinfo{year}{1967}).

\bibitem[{\citenamefont{Charney}(1971)}]{Charney1971}
\bibinfo{author}{\bibfnamefont{J.~G.} \bibnamefont{Charney}},
  \bibinfo{journal}{J. Atmos. Sci.} \textbf{\bibinfo{volume}{28}},
  \bibinfo{pages}{1087} (\bibinfo{year}{1971}).

\bibitem[{\citenamefont{M\'etais et~al.}(1996)\citenamefont{M\'etais, Bartello,
  Garnier, Riley, and Lesieur}}]{Metais1996}
\bibinfo{author}{\bibfnamefont{O.}~\bibnamefont{M\'etais}},
  \bibinfo{author}{\bibfnamefont{P.}~\bibnamefont{Bartello}},
  \bibinfo{author}{\bibfnamefont{E.}~\bibnamefont{Garnier}},
  \bibinfo{author}{\bibfnamefont{J.~J.} \bibnamefont{Riley}}, \bibnamefont{and}
  \bibinfo{author}{\bibfnamefont{M.}~\bibnamefont{Lesieur}},
  \bibinfo{journal}{Dyn. Atmos. Oceans} \textbf{\bibinfo{volume}{23}},
  \bibinfo{pages}{193} (\bibinfo{year}{1996}),
  \urlprefix\url{http://www.sciencedirect.com/science/article/pii/0377026595004130}.

\bibitem[{\citenamefont{Smith and Waleffe}(2002)}]{LSmith2002}
\bibinfo{author}{\bibfnamefont{L.~M.} \bibnamefont{Smith}} \bibnamefont{and}
  \bibinfo{author}{\bibfnamefont{F.}~\bibnamefont{Waleffe}},
  \bibinfo{journal}{J. Fluid Mech.} \textbf{\bibinfo{volume}{451}},
  \bibinfo{pages}{145} (\bibinfo{year}{2002}),
  \urlprefix\url{http://www.journals.cambridge.org/abstract_S0022112001006309}.

\bibitem[{\citenamefont{Kurien et~al.}(2008)\citenamefont{Kurien, Wingate, and
  Taylor}}]{Kurien2008}
\bibinfo{author}{\bibfnamefont{S.}~\bibnamefont{Kurien}},
  \bibinfo{author}{\bibfnamefont{B.}~\bibnamefont{Wingate}}, \bibnamefont{and}
  \bibinfo{author}{\bibfnamefont{M.~A.} \bibnamefont{Taylor}},
  \bibinfo{journal}{Europhys. Lett.} \textbf{\bibinfo{volume}{84}},
  \bibinfo{pages}{24003} (\bibinfo{year}{2008}),
  \urlprefix\url{http://stacks.iop.org/0295-5075/84/i=2/a=24003?key=crossref.115925eb0eca1c5328d37bcf2ab88bf0}.

\bibitem[{\citenamefont{Marino et~al.}(2013)\citenamefont{Marino, Mininni,
  Rosenberg, and Pouquet}}]{Marino2013b}
\bibinfo{author}{\bibfnamefont{R.}~\bibnamefont{Marino}},
  \bibinfo{author}{\bibfnamefont{P.~D.} \bibnamefont{Mininni}},
  \bibinfo{author}{\bibfnamefont{D.}~\bibnamefont{Rosenberg}},
  \bibnamefont{and} \bibinfo{author}{\bibfnamefont{A.}~\bibnamefont{Pouquet}},
  \bibinfo{journal}{Europhys. Lett.} \textbf{\bibinfo{volume}{102}},
  \bibinfo{pages}{44006} (\bibinfo{year}{2013}).

\bibitem[{\citenamefont{Herbert et~al.}(unpublished)\citenamefont{Herbert,
  Pouquet, and Marino}}]{Herbert2014b}
\bibinfo{author}{\bibfnamefont{C.}~\bibnamefont{Herbert}},
  \bibinfo{author}{\bibfnamefont{A.}~\bibnamefont{Pouquet}}, \bibnamefont{and}
  \bibinfo{author}{\bibfnamefont{R.}~\bibnamefont{Marino}},
  \bibinfo{note}{{arXiv:1401.2103}},
  \urlprefix\url{http://arxiv.org/abs/1401.2103}.

\bibitem[{\citenamefont{Herbert}(2014)}]{Herbert2014a}
\bibinfo{author}{\bibfnamefont{C.}~\bibnamefont{Herbert}},
  \bibinfo{journal}{Phys. Rev. E} \textbf{\bibinfo{volume}{89}},
  \bibinfo{pages}{013010} (\bibinfo{year}{2014}),
  \bibinfo{note}{arXiv:1311.3971},
  \urlprefix\url{http://arxiv.org/abs/1311.3971}.

\bibitem[{\citenamefont{Zhu et~al.}(2014)\citenamefont{Zhu, Yang, and
  Zhu}}]{Zhu2014}
\bibinfo{author}{\bibfnamefont{J.-Z.} \bibnamefont{Zhu}},
  \bibinfo{author}{\bibfnamefont{W.}~\bibnamefont{Yang}}, \bibnamefont{and}
  \bibinfo{author}{\bibfnamefont{G.-Y.} \bibnamefont{Zhu}},
  \bibinfo{journal}{J. Fluid Mech.} \textbf{\bibinfo{volume}{739}},
  \bibinfo{pages}{479} (\bibinfo{year}{2014}),
  \urlprefix\url{http://arxiv.org/abs/1303.3823}.

\bibitem[{\citenamefont{Herbert
  et~al.}(2012{\natexlab{a}})\citenamefont{Herbert, Daviaud, Dubrulle,
  Nazarenko, and Naso}}]{EricHerbert2012}
\bibinfo{author}{\bibfnamefont{E.}~\bibnamefont{Herbert}},
  \bibinfo{author}{\bibfnamefont{F.}~\bibnamefont{Daviaud}},
  \bibinfo{author}{\bibfnamefont{B.}~\bibnamefont{Dubrulle}},
  \bibinfo{author}{\bibfnamefont{S.~V.} \bibnamefont{Nazarenko}},
  \bibnamefont{and} \bibinfo{author}{\bibfnamefont{A.}~\bibnamefont{Naso}},
  \bibinfo{journal}{Europhys. Lett.} \textbf{\bibinfo{volume}{100}},
  \bibinfo{pages}{44003} (\bibinfo{year}{2012}{\natexlab{a}}),
  \urlprefix\url{http://stacks.iop.org/0295-5075/100/i=4/a=44003?key=crossref.6a482ce788e7445b50b725c2f2134b1a}.

\bibitem[{\citenamefont{Biferale et~al.}(2012)\citenamefont{Biferale,
  Musacchio, and Toschi}}]{Biferale2012}
\bibinfo{author}{\bibfnamefont{L.}~\bibnamefont{Biferale}},
  \bibinfo{author}{\bibfnamefont{S.}~\bibnamefont{Musacchio}},
  \bibnamefont{and} \bibinfo{author}{\bibfnamefont{F.}~\bibnamefont{Toschi}},
  \bibinfo{journal}{Phys. Rev. Lett.} \textbf{\bibinfo{volume}{108}},
  \bibinfo{pages}{164501} (\bibinfo{year}{2012}),
  \urlprefix\url{http://link.aps.org/doi/10.1103/PhysRevLett.108.164501}.

\bibitem[{\citenamefont{Lucarini et~al.}(unpublished)\citenamefont{Lucarini,
  Blender, Herbert, Pascale, Ragone, and Wouters}}]{Lucarini2013}
\bibinfo{author}{\bibfnamefont{V.}~\bibnamefont{Lucarini}},
  \bibinfo{author}{\bibfnamefont{R.}~\bibnamefont{Blender}},
  \bibinfo{author}{\bibfnamefont{C.}~\bibnamefont{Herbert}},
  \bibinfo{author}{\bibfnamefont{S.}~\bibnamefont{Pascale}},
  \bibinfo{author}{\bibfnamefont{F.}~\bibnamefont{Ragone}}, \bibnamefont{and}
  \bibinfo{author}{\bibfnamefont{J.}~\bibnamefont{Wouters}},
  \bibinfo{note}{arXiv:1311.1190},
  \urlprefix\url{http://arxiv.org/abs/1311.1190}.

\bibitem[{\citenamefont{Lee}(1952)}]{Lee1952}
\bibinfo{author}{\bibfnamefont{T.~D.} \bibnamefont{Lee}}, \bibinfo{journal}{Q.
  Appl. Math.} \textbf{\bibinfo{volume}{10}}, \bibinfo{pages}{69}
  (\bibinfo{year}{1952}).

\bibitem[{\citenamefont{Kraichnan}(1975)}]{Kraichnan1975}
\bibinfo{author}{\bibfnamefont{R.~H.} \bibnamefont{Kraichnan}},
  \bibinfo{journal}{J. Fluid Mech.} \textbf{\bibinfo{volume}{67}},
  \bibinfo{pages}{155} (\bibinfo{year}{1975}).

\bibitem[{\citenamefont{Kraichnan and Montgomery}(1980)}]{Kraichnan1980}
\bibinfo{author}{\bibfnamefont{R.~H.} \bibnamefont{Kraichnan}}
  \bibnamefont{and} \bibinfo{author}{\bibfnamefont{D.~C.}
  \bibnamefont{Montgomery}}, \bibinfo{journal}{Rep. Prog. Phys.}
  \textbf{\bibinfo{volume}{43}}, \bibinfo{pages}{547} (\bibinfo{year}{1980}).

\bibitem[{\citenamefont{Miller}(1990)}]{Miller1990}
\bibinfo{author}{\bibfnamefont{J.}~\bibnamefont{Miller}},
  \bibinfo{journal}{Phys. Rev. Lett.} \textbf{\bibinfo{volume}{65}},
  \bibinfo{pages}{2137} (\bibinfo{year}{1990}).

\bibitem[{\citenamefont{Robert and Sommeria}(1991)}]{Robert1991a}
\bibinfo{author}{\bibfnamefont{R.}~\bibnamefont{Robert}} \bibnamefont{and}
  \bibinfo{author}{\bibfnamefont{J.}~\bibnamefont{Sommeria}},
  \bibinfo{journal}{J. Fluid Mech.} \textbf{\bibinfo{volume}{229}},
  \bibinfo{pages}{291} (\bibinfo{year}{1991}).

\bibitem[{\citenamefont{Robert}(1991)}]{Robert1991b}
\bibinfo{author}{\bibfnamefont{R.}~\bibnamefont{Robert}}, \bibinfo{journal}{J.
  Stat. Phys.} \textbf{\bibinfo{volume}{65}}, \bibinfo{pages}{531}
  (\bibinfo{year}{1991}).

\bibitem[{\citenamefont{Chavanis and Sommeria}(1996)}]{Chavanis1996a}
\bibinfo{author}{\bibfnamefont{P.-H.} \bibnamefont{Chavanis}} \bibnamefont{and}
  \bibinfo{author}{\bibfnamefont{J.}~\bibnamefont{Sommeria}},
  \bibinfo{journal}{J. Fluid Mech.} \textbf{\bibinfo{volume}{314}},
  \bibinfo{pages}{267} (\bibinfo{year}{1996}).

\bibitem[{\citenamefont{Chen and Cross}(1997)}]{Chen1997}
\bibinfo{author}{\bibfnamefont{P.}~\bibnamefont{Chen}} \bibnamefont{and}
  \bibinfo{author}{\bibfnamefont{M.~C.} \bibnamefont{Cross}},
  \bibinfo{journal}{Phys. Rev. E} \textbf{\bibinfo{volume}{56}},
  \bibinfo{pages}{2284} (\bibinfo{year}{1997}).

\bibitem[{\citenamefont{Chavanis and Sommeria}(1998)}]{Chavanis1998b}
\bibinfo{author}{\bibfnamefont{P.-H.} \bibnamefont{Chavanis}} \bibnamefont{and}
  \bibinfo{author}{\bibfnamefont{J.}~\bibnamefont{Sommeria}},
  \bibinfo{journal}{J. Fluid Mech.} \textbf{\bibinfo{volume}{356}},
  \bibinfo{pages}{259} (\bibinfo{year}{1998}),
  \urlprefix\url{http://journals.cambridge.org/abstract_S0022112097007933}.

\bibitem[{\citenamefont{Venaille and Bouchet}(2009)}]{Venaille2009}
\bibinfo{author}{\bibfnamefont{A.}~\bibnamefont{Venaille}} \bibnamefont{and}
  \bibinfo{author}{\bibfnamefont{F.}~\bibnamefont{Bouchet}},
  \bibinfo{journal}{Phys. Rev. Lett.} \textbf{\bibinfo{volume}{102}},
  \bibinfo{pages}{104501} (\bibinfo{year}{2009}).

\bibitem[{\citenamefont{Herbert
  et~al.}(2012{\natexlab{b}})\citenamefont{Herbert, Dubrulle, Chavanis, and
  Paillard}}]{Herbert2012a}
\bibinfo{author}{\bibfnamefont{C.}~\bibnamefont{Herbert}},
  \bibinfo{author}{\bibfnamefont{B.}~\bibnamefont{Dubrulle}},
  \bibinfo{author}{\bibfnamefont{P.-H.} \bibnamefont{Chavanis}},
  \bibnamefont{and} \bibinfo{author}{\bibfnamefont{D.}~\bibnamefont{Paillard}},
  \bibinfo{journal}{Phys. Rev. E} \textbf{\bibinfo{volume}{85}},
  \bibinfo{pages}{056304} (\bibinfo{year}{2012}{\natexlab{b}}).

\bibitem[{\citenamefont{Bouchet and Sommeria}(2002)}]{Bouchet2002}
\bibinfo{author}{\bibfnamefont{F.}~\bibnamefont{Bouchet}} \bibnamefont{and}
  \bibinfo{author}{\bibfnamefont{J.}~\bibnamefont{Sommeria}},
  \bibinfo{journal}{J. Fluid Mech.} \textbf{\bibinfo{volume}{464}},
  \bibinfo{pages}{165} (\bibinfo{year}{2002}).

\bibitem[{\citenamefont{Naso et~al.}(2011)\citenamefont{Naso, Chavanis, and
  Dubrulle}}]{Naso2011}
\bibinfo{author}{\bibfnamefont{A.}~\bibnamefont{Naso}},
  \bibinfo{author}{\bibfnamefont{P.-H.} \bibnamefont{Chavanis}},
  \bibnamefont{and} \bibinfo{author}{\bibfnamefont{B.}~\bibnamefont{Dubrulle}},
  \bibinfo{journal}{Eur. Phys. J. B} \textbf{\bibinfo{volume}{80}},
  \bibinfo{pages}{493} (\bibinfo{year}{2011}).

\bibitem[{\citenamefont{Venaille and Bouchet}(2011)}]{Venaille2011a}
\bibinfo{author}{\bibfnamefont{A.}~\bibnamefont{Venaille}} \bibnamefont{and}
  \bibinfo{author}{\bibfnamefont{F.}~\bibnamefont{Bouchet}},
  \bibinfo{journal}{J. Stat. Phys.} \textbf{\bibinfo{volume}{143}},
  \bibinfo{pages}{346} (\bibinfo{year}{2011}).

\bibitem[{\citenamefont{Herbert
  et~al.}(2012{\natexlab{c}})\citenamefont{Herbert, Dubrulle, Chavanis, and
  Paillard}}]{Herbert2012b}
\bibinfo{author}{\bibfnamefont{C.}~\bibnamefont{Herbert}},
  \bibinfo{author}{\bibfnamefont{B.}~\bibnamefont{Dubrulle}},
  \bibinfo{author}{\bibfnamefont{P.-H.} \bibnamefont{Chavanis}},
  \bibnamefont{and} \bibinfo{author}{\bibfnamefont{D.}~\bibnamefont{Paillard}},
  \bibinfo{journal}{J. Stat. Mech.} \textbf{\bibinfo{volume}{2012}},
  \bibinfo{pages}{P05023} (\bibinfo{year}{2012}{\natexlab{c}}).

\bibitem[{\citenamefont{Salmon et~al.}(1976)\citenamefont{Salmon, Holloway, and
  Hendershott}}]{Salmon1976}
\bibinfo{author}{\bibfnamefont{R.}~\bibnamefont{Salmon}},
  \bibinfo{author}{\bibfnamefont{G.}~\bibnamefont{Holloway}}, \bibnamefont{and}
  \bibinfo{author}{\bibfnamefont{M.~C.} \bibnamefont{Hendershott}},
  \bibinfo{journal}{J. Fluid Mech.} \textbf{\bibinfo{volume}{75}},
  \bibinfo{pages}{691} (\bibinfo{year}{1976}).

\bibitem[{\citenamefont{Merryfield}(1998)}]{Merryfield1998}
\bibinfo{author}{\bibfnamefont{W.~J.} \bibnamefont{Merryfield}},
  \bibinfo{journal}{J. Fluid Mech.} \textbf{\bibinfo{volume}{354}},
  \bibinfo{pages}{345} (\bibinfo{year}{1998}),
  \urlprefix\url{http://journals.cambridge.org/abstract_S0022112097007684}.

\bibitem[{\citenamefont{Venaille}(2012)}]{Venaille2012b}
\bibinfo{author}{\bibfnamefont{A.}~\bibnamefont{Venaille}},
  \bibinfo{journal}{J. Fluid Mech.} \textbf{\bibinfo{volume}{699}},
  \bibinfo{pages}{500} (\bibinfo{year}{2012}).

\bibitem[{\citenamefont{Venaille et~al.}(2012)\citenamefont{Venaille, Vallis,
  and Griffies}}]{Venaille2012a}
\bibinfo{author}{\bibfnamefont{A.}~\bibnamefont{Venaille}},
  \bibinfo{author}{\bibfnamefont{G.~K.} \bibnamefont{Vallis}},
  \bibnamefont{and} \bibinfo{author}{\bibfnamefont{S.~M.}
  \bibnamefont{Griffies}}, \bibinfo{journal}{J. Fluid Mech.}
  \textbf{\bibinfo{volume}{709}}, \bibinfo{pages}{490} (\bibinfo{year}{2012}).

\bibitem[{\citenamefont{Pedlosky}(1987)}]{PedloskyGFD}
\bibinfo{author}{\bibfnamefont{J.}~\bibnamefont{Pedlosky}},
  \emph{\bibinfo{title}{Geophysical Fluid Dynamics}}
  (\bibinfo{publisher}{Springer, Berlin}, \bibinfo{year}{1987}).

\bibitem[{\citenamefont{Vallis}(2006)}]{VallisBook}
\bibinfo{author}{\bibfnamefont{G.~K.} \bibnamefont{Vallis}},
  \emph{\bibinfo{title}{{Atmospheric and Oceanic Fluid Dynamics: Fundamentals
  and Large-scale Circulation}}} (\bibinfo{publisher}{Cambridge University
  Press}, \bibinfo{year}{2006}).

\bibitem[{\citenamefont{Herbert}(2013)}]{Herbert2013b}
\bibinfo{author}{\bibfnamefont{C.}~\bibnamefont{Herbert}}, \bibinfo{journal}{J.
  Stat. Phys.} \textbf{\bibinfo{volume}{152}}, \bibinfo{pages}{1084}
  (\bibinfo{year}{2013}).

\bibitem[{\citenamefont{Michel and Robert}(1994)}]{Michel1994b}
\bibinfo{author}{\bibfnamefont{J.}~\bibnamefont{Michel}} \bibnamefont{and}
  \bibinfo{author}{\bibfnamefont{R.}~\bibnamefont{Robert}},
  \bibinfo{journal}{Commun. Math. Phys.} \textbf{\bibinfo{volume}{159}},
  \bibinfo{pages}{195} (\bibinfo{year}{1994}).

\bibitem[{\citenamefont{Robert}(2000)}]{Robert2000}
\bibinfo{author}{\bibfnamefont{R.}~\bibnamefont{Robert}},
  \bibinfo{journal}{Commun. Math. Phys.} \textbf{\bibinfo{volume}{212}},
  \bibinfo{pages}{245} (\bibinfo{year}{2000}).

\bibitem[{\citenamefont{Bouchet and Corvellec}(2010)}]{Bouchet2010}
\bibinfo{author}{\bibfnamefont{F.}~\bibnamefont{Bouchet}} \bibnamefont{and}
  \bibinfo{author}{\bibfnamefont{M.}~\bibnamefont{Corvellec}},
  \bibinfo{journal}{J. Stat. Mech.} \textbf{\bibinfo{volume}{2010}},
  \bibinfo{pages}{P08021} (\bibinfo{year}{2010}).

\bibitem[{\citenamefont{Bouchet}(2008)}]{Bouchet2008}
\bibinfo{author}{\bibfnamefont{F.}~\bibnamefont{Bouchet}},
  \bibinfo{journal}{Physica D} \textbf{\bibinfo{volume}{237}},
  \bibinfo{pages}{1976} (\bibinfo{year}{2008}).

\bibitem[{\citenamefont{Chavanis}(2008)}]{Chavanis2008b}
\bibinfo{author}{\bibfnamefont{P.-H.} \bibnamefont{Chavanis}},
  \bibinfo{journal}{Physica D} \textbf{\bibinfo{volume}{237}},
  \bibinfo{pages}{1998} (\bibinfo{year}{2008}).

\bibitem[{\citenamefont{Turkington}(1999)}]{Turkington1999}
\bibinfo{author}{\bibfnamefont{B.}~\bibnamefont{Turkington}},
  \bibinfo{journal}{Comm. Pure Appl. Math.} \textbf{\bibinfo{volume}{52}},
  \bibinfo{pages}{781} (\bibinfo{year}{1999}).

\bibitem[{\citenamefont{Ellis et~al.}(2000)\citenamefont{Ellis, Haven, and
  Turkington}}]{Ellis2000}
\bibinfo{author}{\bibfnamefont{R.~S.} \bibnamefont{Ellis}},
  \bibinfo{author}{\bibfnamefont{K.}~\bibnamefont{Haven}}, \bibnamefont{and}
  \bibinfo{author}{\bibfnamefont{B.}~\bibnamefont{Turkington}},
  \bibinfo{journal}{J. Stat. Phys.} \textbf{\bibinfo{volume}{101}},
  \bibinfo{pages}{999} (\bibinfo{year}{2000}).

\bibitem[{\citenamefont{Ellis et~al.}(2002)\citenamefont{Ellis, Haven, and
  Turkington}}]{Ellis2002}
\bibinfo{author}{\bibfnamefont{R.~S.} \bibnamefont{Ellis}},
  \bibinfo{author}{\bibfnamefont{K.}~\bibnamefont{Haven}}, \bibnamefont{and}
  \bibinfo{author}{\bibfnamefont{B.}~\bibnamefont{Turkington}},
  \bibinfo{journal}{Nonlinearity} \textbf{\bibinfo{volume}{15}},
  \bibinfo{pages}{239} (\bibinfo{year}{2002}).

\bibitem[{\citenamefont{Naso et~al.}(2010)\citenamefont{Naso, Chavanis, and
  Dubrulle}}]{Naso2010a}
\bibinfo{author}{\bibfnamefont{A.}~\bibnamefont{Naso}},
  \bibinfo{author}{\bibfnamefont{P.-H.} \bibnamefont{Chavanis}},
  \bibnamefont{and} \bibinfo{author}{\bibfnamefont{B.}~\bibnamefont{Dubrulle}},
  \bibinfo{journal}{Eur. Phys. J. B} \textbf{\bibinfo{volume}{77}},
  \bibinfo{pages}{187} (\bibinfo{year}{2010}).

\bibitem[{\citenamefont{Chavanis}(2009)}]{Chavanis2009}
\bibinfo{author}{\bibfnamefont{P.-H.} \bibnamefont{Chavanis}},
  \bibinfo{journal}{Eur. Phys. J. B} \textbf{\bibinfo{volume}{70}},
  \bibinfo{pages}{73} (\bibinfo{year}{2009}).

\bibitem[{\citenamefont{Qi and Marston}(unpublished)}]{Qi2014}
\bibinfo{author}{\bibfnamefont{W.}~\bibnamefont{Qi}} \bibnamefont{and}
  \bibinfo{author}{\bibfnamefont{J.~B.} \bibnamefont{Marston}},
  \eprint{1312.2553v1}, \urlprefix\url{http://arxiv.org/abs/1312.2553v1}.

\bibitem[{\citenamefont{Salmon}(1978)}]{Salmon1978}
\bibinfo{author}{\bibfnamefont{R.}~\bibnamefont{Salmon}},
  \bibinfo{journal}{Geophys. Astrophys. Fluid Dyn.}
  \textbf{\bibinfo{volume}{10}}, \bibinfo{pages}{25} (\bibinfo{year}{1978}).

\bibitem[{\citenamefont{Rhines}(1979)}]{Rhines1979}
\bibinfo{author}{\bibfnamefont{P.~B.} \bibnamefont{Rhines}},
  \bibinfo{journal}{Ann. Rev. Fluid Mech.} \textbf{\bibinfo{volume}{11}},
  \bibinfo{pages}{401} (\bibinfo{year}{1979}).

\bibitem[{\citenamefont{Salmon}(1980)}]{Salmon1980}
\bibinfo{author}{\bibfnamefont{R.}~\bibnamefont{Salmon}},
  \bibinfo{journal}{Geophys. Astrophys. Fluid Dyn.}
  \textbf{\bibinfo{volume}{15}}, \bibinfo{pages}{167} (\bibinfo{year}{1980}).

\bibitem[{\citenamefont{Rhines}(1975)}]{Rhines1975}
\bibinfo{author}{\bibfnamefont{P.~B.} \bibnamefont{Rhines}},
  \bibinfo{journal}{J. Fluid Mech.} \textbf{\bibinfo{volume}{69}},
  \bibinfo{pages}{417} (\bibinfo{year}{1975}).

\bibitem[{\citenamefont{Vallis and Maltrud}(1993)}]{Vallis1993}
\bibinfo{author}{\bibfnamefont{G.~K.} \bibnamefont{Vallis}} \bibnamefont{and}
  \bibinfo{author}{\bibfnamefont{M.~E.} \bibnamefont{Maltrud}},
  \bibinfo{journal}{J. Phys. Oceanogr.} \textbf{\bibinfo{volume}{23}},
  \bibinfo{pages}{1346} (\bibinfo{year}{1993}).

\end{thebibliography}
\end{document}